\documentclass{aa}  
\usepackage{graphicx}
\usepackage{graphicx,xcolor}
\usepackage{txfonts}
\usepackage{multirow}
\usepackage{lscape}
\usepackage{longtable}
\usepackage[breaklinks=true]{hyperref}
    \hypersetup{
        colorlinks=true,
        linkcolor=blue,
        filecolor=magenta,      
        urlcolor=blue,
        citecolor=blue
    }
\usepackage{tabularx}

\begin{document}

   \title{Updated extraction of the APOGEE \texorpdfstring{1.5273 $\mu m$}{} diffuse interstellar band:
a Planck view\thanks{Based on SDSS/APOGEE Archive data and observations obtained with Planck (http://www.esa.int/Planck), an ESA science mission with instruments and contributions directly funded by ESA Member States, NASA, and Canada} on the carrier depletion in dense cores
%The skin effect of the \texorpdfstring{1.5273 $\mu m$}{} Near-Infrared Diffuse Interstellar Band: a Planck 
}

   %\subtitle{I. Overviewing the edge effect -mechanism}
\titlerunning{Updated extraction of the 1.5273 $\mu$ DIB: a Planck view}
   \author{M. Elyajouri
          \inst{1}
          \and
          R. Lallement\inst{1}
          }

   \institute{GEPI, Observatoire de Paris, PSL University, CNRS,
Place Jules Janssen, 92190 Meudon, France\\
              \email{meriem.el-yajouri@obspm.fr}
             }

   \date{Received ; accepted }

  \abstract
   {Constraining the spatial distribution of Diffuse Interstellar Band (DIB) carriers and their links with gas and dust are mandatory steps  in understanding their role in interstellar chemistry.}
  % aims heading (mandatory)
   {The latest SDSS/APOGEE data release DR14 has provided an increased number of stellar spectra in the H band and  associated stellar models using an innovative algorithm known as The \emph{Cannon}. We took advantage of these novelties to extract the 15\,273 $\mbox{\AA}$ near-infrared DIB and to study its link with dust extinction and emission.
   }
  % methods heading (mandatory)
   {We modified our automated fitting methods dedicated for hot stars and used in earlier studies with some adaptations motivated by the change from early- or intermediate-type stars to red giants. A new method has also been developed to quantify the upper limits on DIB strengths. Careful and thorough examinations of the DIB parameters, the continuum shape and the quality of the adjustment of the model to the data were done. We compared our DIB measurements with the stellar extinctions A$_V$ from the Starhorse database. We then compared the resulting DIB-extinction ratio with the dust optical depth derived from Planck data, globally and also separately for nearby off-Plane cloud complexes.
   }
  % results heading (mandatory)
   {Our analysis has led to the production of a catalog containing 124\,064 new measurements of the 15\,273 $\mbox{\AA}$ DIB, allowing us to revisit the correlation between DIB strength and dust reddening.  The new data reveal clearly that the sky-averaged 15\,273 $\mbox{\AA}$ DIB strength is linearly correlated with A$_V$ over two orders as reported by earlier studies but leveling off with respect to extinction for highly reddened lines-of sight behind dense clouds. The comparison with Planck individual optical depths reveals in a conspicuous way this DIB depletion in the dense cores and shows it applies to all off-Plane dense clouds. Using selected targets located beyond the Orion, Taurus and Cepheus clouds, we derived empirical relationships between the DIB to extinction ratio and the Planck dust optical depth for the three cloud complexes. Their average is similar to the DIB carrier depletion measured in the dark cloud Barnard 68.}
  % conclusions heading (optional), leave it empty if necessary 
   {APOGEE measurements confirm the ubiquity of the 15\,273 $\mbox{\AA}$ DIB carrier decrease with respect to dust grains in dense cloud cores, in a manner that can be empirically related to the dust optical depth reached in the cloud. They also show that the ratio between the DIB equivalent width and the extinction A$_{V}$ for sightlines with $\tau$(353GHz)  $\lesssim$ 2 $\times 10^{-5}$ that do not contain dense molecular gas is about four times higher than the constant limit towards which the ratio tends for very long sightlines with many diffuse and dense phases distributed in distance.
   }

   \maketitle

\section{Introduction}

One of the most challenging problems in terms of InterStellar Medium (ISM) composition and evolution is the ubiquitous presence of the numerous, unidentified Diffuse Interstellar Bands (DIBs) detected in our Galaxy and other galaxies. Since their first discoveries in the 1920's by \citet{Heger22}, these absorption features seen in spectra of Milky Way reddened stars or in galaxy spectra, have puzzled astronomers and spectroscopists (see \citet{McCall13}, for a recent historical review). 
In spite of an apparent general correlation between the DIB strengths and the interstellar extinction or reddening, all efforts have failed up to confirm the hypothesis that dust particles are responsible for the absorption of DIBs. 
One can cite in particular the most stringent limitations on polarization reported by \cite{cox2007a} who analyzed six strong DIBs (at 5780, 5797, 6196, 6284, 6379 and 6613 $\AA$) and revealed no linear polarization, a polarization expected for large grain-related carriers. The lack of correlation between DIB strengths and the Far-UV (FUV) extinction disfavors the small grain-related carriers as well \citep{Desert1995, Xiang2017}.
DIB carriers are then believed to be large molecules in gas phase. The detection of substructures in the profile of some DIBs \citep[e.g.][and references therein]{Elyajouri2018} supports the molecular nature of some DIB carriers. 

The most promising carrier candidates are thought to be carbon chains, polycyclic aromatic hydrocarbons (PAHs), and fullerenes. Indeed, at least four of five bands predicted to be associated with the buckminsterfullerene C$_{60}^{+}$ have now been detected, making this fullerene cation the first identified DIB carrier \citep{Foing94,Cordiner17,Campbell16,Lallement2018a}.

It remains that the global correlation between most DIBs and optical extinction A$_{V}$ suggests that dust grains that cause this extinction are associated with/linked to DIB carriers in some way. However, the underlying mechanisms and role of DIB carriers in the multiple physico-chemical reaction chains that occur throughout the ISM life cycle are still unknown and many open questions remain to be answered: is the DIB carrier density related to the depletion onto the grains? can grains be the formation site of DIB carriers? 

Fortunately, the new massive stellar spectroscopic surveys can now support statistical studies allowing comparisons between absorption by gaseous species (e.g. CaII, NaI, KI, C$_{2}$, CH, CH$^{+}$), DIB carriers and dust that can help clarify their links. Such statistical studies of relationships between DIBs, gas and dust grains as well as ratios between pairs of diffuse bands contain precious information on the molecular content and the evolution of macro-molecules and grains in response to the physical properties, the radiation field, the abundances in the ISM, and the grain size distribution. A hierarchy of DIBs ordered by increasing sensitivity to the ionizing field is in progress \citep[e.g.][]{Ensor2017}. The link with the ionizing field is particularly interesting as it is related to the depth within the clouds, and thus to the grain properties. 

An important observational clue is that some of the strongest optical DIBs have shown to level off with respect to dust reddening for lines-of-sight that probe denser molecular cores, rather than the diffuse edges, of interstellar clouds. Using the Lick Observatory scanning spectrophotometer, \citet{Wampler1966} was the first to notice such an effect for the DIB at 4430 $\mbox{\AA}$ in the spectra of background dark clouds stars. It has been also seen by \citet{Snow1974} basing it on their study of DIB strengths at 4430, 5780, and 5797 $\mbox{\AA}$ in Sco-Oph and Per. They concluded that formation or survival of the carrier is inhibited in cloud interiors. Since then, the same effect has been commented in a number of studies \citep{Strom1975,Meyer1984,Wallerstein1987, Adamson1991, Herbig95}. This behaviour has been referred to as the skin effect in the \cite{Herbig95} review. It is supported by the higher degree of correlation of the DIB strength with HI compared to molecular hydrogen \citep{Friedman11}. Moreover, the discovery of unusually weak DIBs in the spectra of the star HD 62542 by \cite{Snow2002} also confirmed such an interpretation, if, as it is likely, the diffuse outer layers of the dark cloud in front of the star have been stripped away. However, a particular class of DIBs, the so-called C2-DIBs \citep{Thorburn03}, seems to be abundant in the dense molecular phase \citep{Adamk05}. For recent reviews including the skin effect see  \cite{SnowMcCall2006}
and \cite{Snow2014}.

More recently, \citet{Lan2015} reported this behavior for many DIBs towards lines of sight intercepting high latitude molecular clouds using the Sloan Digital Sky Survey (SDSS) star, galaxy and quasar spectra. The authors concluded that the break seen at high E(B$-$V) values can be quantitatively characterized by the decrease of \ion{HI} in dense clouds, and higher molecular fraction.

Finally, important additional findings on the skin effect have been brought by \cite{Fan2017}. The authors have performed a large study of eight different, strong optical DIBs and, especially, produced a statistically significant study of the DIB to extinction ratio as a function of various atomic and molecular tracers of the ISM phase. In particular, they found that the relationship between the DIB to extinction ratio and the molecular hydrogen fraction f(H$_{2}$) is characterized by a  \emph{lambda-shaped} curve with a peak at  f(H$_{2}$)$\simeq$0.3, definitely showing that the DIB carriers reside predominantly in this range of molecular fraction and tend to decrease in both fully molecular cores and atomic gas. As a result, this study reveals how the physical state of the gas controls the DIB to extinction ratio. Here, the term \emph{skin effect} takes all this sense, since there is indeed a maximum volume density of DIB carriers for a particular molecular fraction, and a decrease above and below this specific fraction value. In what follows, we will only essentially address the depletion in the dense molecular phase, and for this reason we will refer to a depletion effect instead of a \emph{skin effet.} 

At least one of the near-infrared (NIR) DIBs, which allow to probe more heavily obscured regions, and thus denser (higher extinction) clouds, also displays a depletion in the dense phase. Using VLT/KMOS moderate resolution (R = 4000) near-infrared (H-band) spectra of 85 stars located behind the Barnard 68 dark globule -- from the edge to the centre, a tomography of the 15\,273 $\mbox{\AA}$ near-infrared DIB in an individual dark interstellar cloud has been performed for the first time \citep{Elyajouri2017b}. It has demonstrated the decrease of the 15\,273 DIB carrier in the denser parts of such dark clouds.

Previous efforts to observe the skin effect either cover small regions around individual clouds, or have insufficient sample sizes to statistically constrain its amplitude.  Taking advantage of the increasing number of stellar surveys with high multiplex instruments, we developed an entirely automated fitting method to extract as much as possible information from stellar spectra. Such methods have been applied to NIR spectra of hot stars, in particular, the telluric calibrators of the SDSS/ \emph{Apache Point Observatory Galactic Evolution Experiment} survey in the H band \citep[][hereafter EMRL16]{Elyajouri16}.  
\defcitealias{Elyajouri16}{EMRL16}	
In this work, we expand the studies to the latest data release DR14 using spectra of red giant stars and covering a significant fraction of the sky. 
Our approach to the problem of the DIB-extinction link involves the use of over one hundred thousand individual lines of sight, focusing exclusively on the strongest DIB in the H-band, at 15\,273 $\mbox{\AA}$. 
We begin the next Sec.~\ref{data} with a description of the public catalogs used in this study, including a brief discussion of the changes and novelties made in the various surveys. 
Sect.~\ref{extraction} describes the methods used for DIB extraction and the resulting catalog. This catalog  will be discussed in detail in separate work, with special focus on its applications.
In Sec.~\ref{correl_ext} we  revisited the relationship between DIB strength and extinction showing a break at increasing extinction values. A special attention is paid to slight differences in the stellar models of the various pipelines, since correction for this variation is crucial to achieve a good DIB extraction, and we discuss the extinction estimates.
Using our results in combination with existing extinction estimates and all-sky dust map, we investigate in Sect.~\ref{skin_effect} both the relationship between their ratio and the dust optical measured by Planck. Our main conclusions are summarized in Sect.~\ref{discussion}.

\section{Description of the public catalogs used in this study} \label{data}

\subsection{SDSS/APOGEE spectra}
The infrared high resolution (R $\sim$ 22\:000) Apache Point Observatory Galactic Evolution Experiment \citep[APOGEE,][]{Majewski2017} is one of the programs in the Sloan Digital Sky Survey III \citep[SDSS-III,][]{Eisenstein11}. Released APOGEE spectra  have a high signal-to-noise (generally above 100) and are remarkably well corrected for telluric emission and absorption with residuals generally below 1\% as can be seen in figures from \cite{Casey2016}. 
The latest APOGEE Data Release 14  \citep[DR14;][]{Abolfathi2018} includes results from APOGEE-1 (September 2011-July 2014) in addition to two years of SDSS-IV APOGEE-2 data (July 2014-July 2016). DR14 APOGEE products contain data from approximately 263\,000 stars, all identified by a 2MASS catalog name.
Among the APOGEE stars, there are 231\,000 main scientific targets, mostly
red giants intended for various scientific programs and forming part of the Bulge, the Bar, the Disk and the Halo.
There are also 27\, 000 young and hot stars called Telluric Standards Stars (TSSs) which were used to correct telluric absorptions \citep{Zasowski13}. These hot stars are distributed in all the fields observed by APOGEE (35 stars/field) and were chosen among the bluest in each field. 17\,000 of these stars from APOGEE Data Release 12 \citep[DR12;][]{Holtzman15} were the subject of a detailed DIB study in \citet{Elyajouri16,Elyajouri2017a}.
They are particularly interesting because one can easily identify any interstellar signature \footnote{The new TSS spectra of DR14 are not publicly available}.
It is obviously more difficult to extract the DIB from the late-type stars since the interstellar features are blended with deep stellar lines. This explains why most of our reliable DIB-fitting results of earlier studies were obtained for hotter stars. However, if the stellar models well reproduce the data, the extraction can be performed without ambiguities due to stellar blends. This is the case of the synthetic spectra of the APOGEE DR14 which includes an innovative data-driven technique, the so-called \emph{Cannon}.
\begin{figure*}
\centering
   \includegraphics[width=\textwidth]{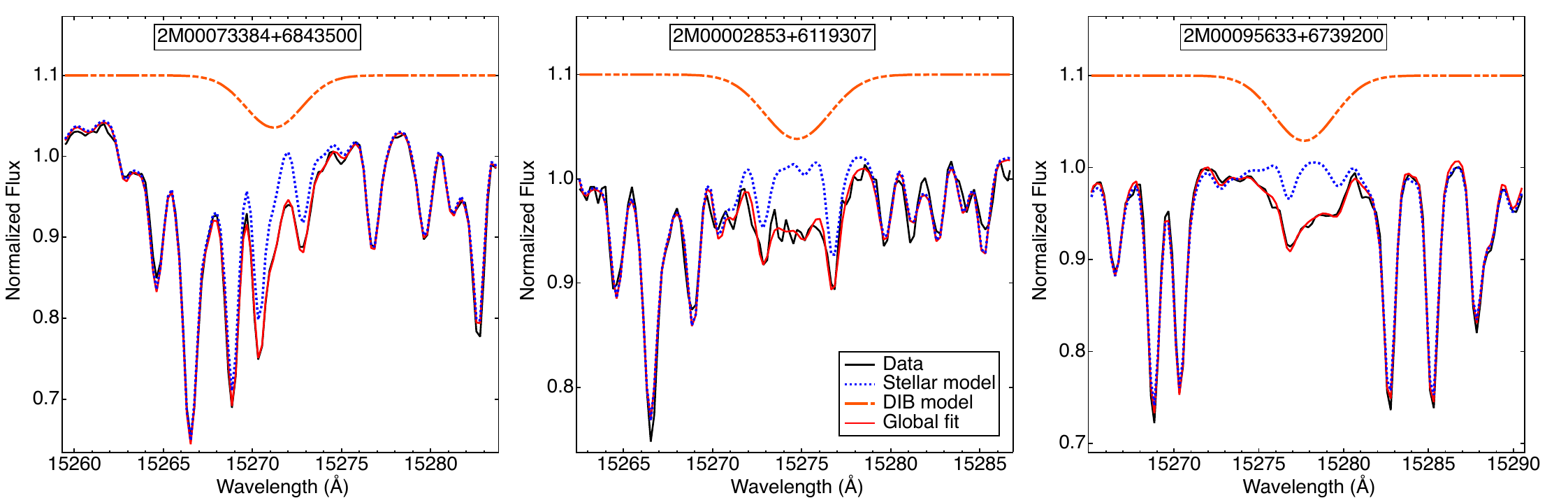}   
    \caption{Illustration of DIB extraction towards three red giants stars. In each figure: APOGEE spectra are shown with solid black curves. The initial stellar model provided by APOGEE \emph{Cannon} is shown in solid blue. The DIB absorption alone is represented in orange (with an offset). The solid magenta curves represent the final stellar+DIB modeled spectrum.
   } \label{fig:my_dr14}
\end{figure*}
\subsection{\emph{Cannon} model spectra}
Highly detailed model spectra were fitted to the APOGEE data and made publicly available for a large number of targets. These fitted models greatly facilitate the extraction of the diffuse interstellar bands present in the APOGEE spectral intervals.
The DR12 synthetic spectra used the APOGEE Stellar Parameter and Chemical Abundances Pipeline \citep[ASPCAP;][]{Garcia2016} for the derivation of the stellar atmospheric parameters and abundances. In DR14 a new and important change is the introduction of an innovative data-driven machine-learning technique known as \emph{Cannon} \citep{Ness2015}. This algorithm parameterized the spectral fluxes as a function of a set of independently-determined stellar parameters and abundances. For DR14, \emph{Cannon}-2 code by \citet{Casey2016}  was trained on ASPCAP products for a subset of high S/N giant stars, and the model was applied to all objects within the range of parameters covered by the training set with some modifications listed in \citet{Abolfathi2018}.

Here we analyze  all available red giant stars (160\, 000) from the public \emph{Cannon} data products. To do so we used the \emph{Cannon}Star file which contain \emph{Cannon} labels, normalized observed spectra and associated uncertainties as well as synthetic spectra for each single star. Such \emph{Cannon} results can be found at path\footnote{\url{http://www.sdss.org/sas/dr14/apogee/spectro/redux/r8/stars/l31c/l31c.2}}/cannon/LOCATION\_ID/cannonStar*.fits. The difference between \emph{Cannon} and ASPCAP has significant consequences on our DIB extraction, as will be detailed in the next section.

\subsection{Starhorse extinction and distance database}
In order to study the link between the 15\,273 $\mbox{\AA}$\: DIB strength and the color excess (section \ref{correl_ext}), we made use of the stellar extinctions in the visible range A$_\mathrm{V}$ calculated with the Bayesian Starhorse method developed by \citet{Santiago2016} and \citet{Queiroz2018}.  The authors applied the Starhorse code to APOGEE DR 14 data and the associated stellar parameters derived from the \emph{Cannon} method, therefore there is a consistency in using the results of the StarHorse technique (here version 1) applied to APOGEE DR14 and the \emph{Cannon} DR14 model spectra. 

Starhorse also contains spectrophotometric distances or parallax distances from the Tycho-Gaia Astrometric Solution \citep[TGAS,][]{Gaia2016} when available, as described in \citet{Anders2018}. We also made use of these distances (see section \ref{skin_effect}). All catalogs are available to the community via the LineA web page\footnote{\url{http://www.linea.gov.br/020-data-center/acesso-a-dados-3/}}. 

\subsection{Planck Galactic dust data}
To trace Galactic dust properties in regions with DIB measurements and allow better comparisons with the interstellar dust along the sightlines, we made use of the Planck Legacy archival \footnote{\url{https://pla.esac.esa.int/pla/}} data. 
The Planck Collaboration used
Planck and WMAP data
to model the spectral energy distributions of the dust
thermal emission and derive an all-sky map of the dust optical
depth at 353 GHz, $\tau_{353}$ and of the dust temperature, $T_\mathrm{D}$. Here we used the Planck 2015 results described in \cite{Planck2016}. 
\section{DIB extraction} \label{extraction}
\begin{figure*}
    \includegraphics[width=0.5\textwidth]{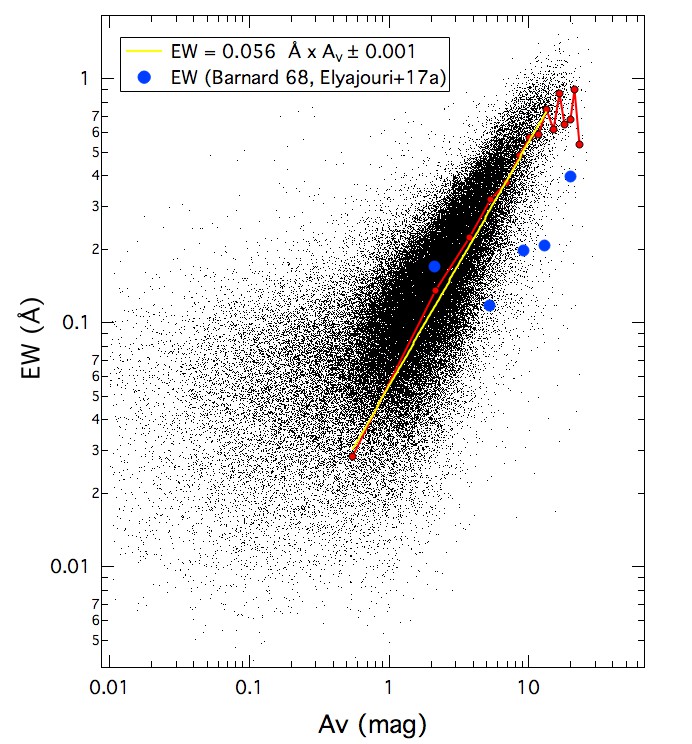}
        \includegraphics[width=0.5\textwidth]{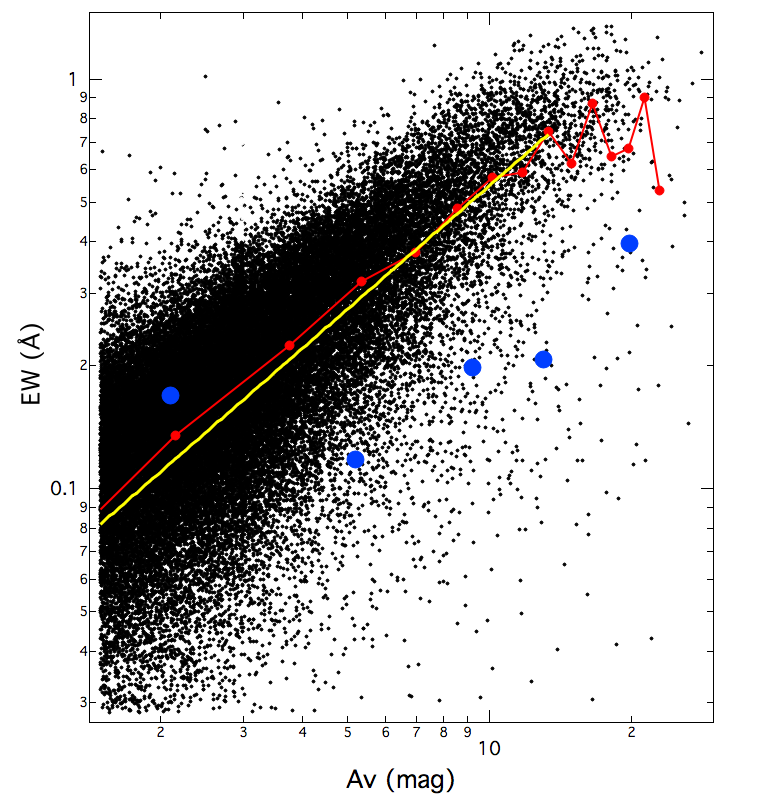}
    \caption{The equivalent width EW of the 15\,273 \AA\ DIB as a function of stellar extinction $A_\mathrm{V}$. Black points are the individual sightline measurements. The red circles are the EW medians in each $\Delta \mathrm{A_V}$ = 1.6 mag, and the yellow line is the fitted EW/$A_\mathrm{V}$ relationship. Superposed are the background stars of Barnard 68 (the larger blue circles).}
    \label{fig:ew_av}
\end{figure*}
\begin{figure*}[h]
    \centering
    \includegraphics[width=0.8\textwidth]{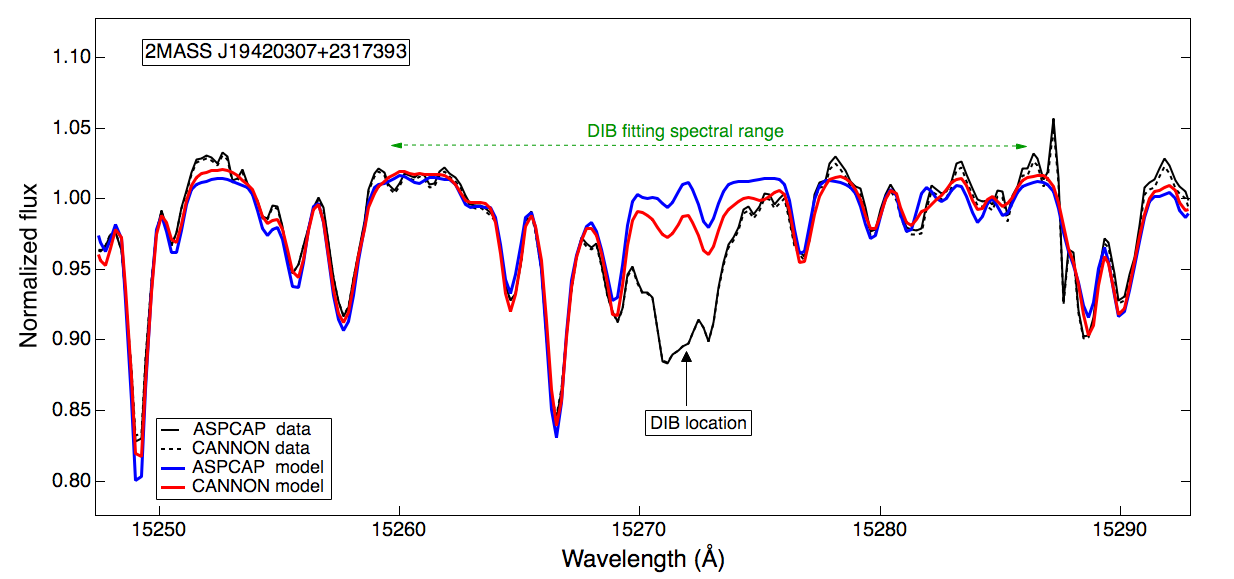}
    \caption{Spectrum of 2MASS J19420307+2317393 from ASPCAP (solid black) and \emph{Cannon}  (dashed line)data products. The associated synthetic stellar models are shown in dashed curves. We clearly see the difference between ASPCAP (blue) and \emph{Cannon} (red) models around the 15\,273 \AA\ DIB location (black arrow). An offset was applied to ASPCAP data and model to facilitate the comparison.}
    \label{fig:my_aspacap_cannon}
\end{figure*}

\begin{figure}[h]
      \includegraphics[width=0.5\textwidth]{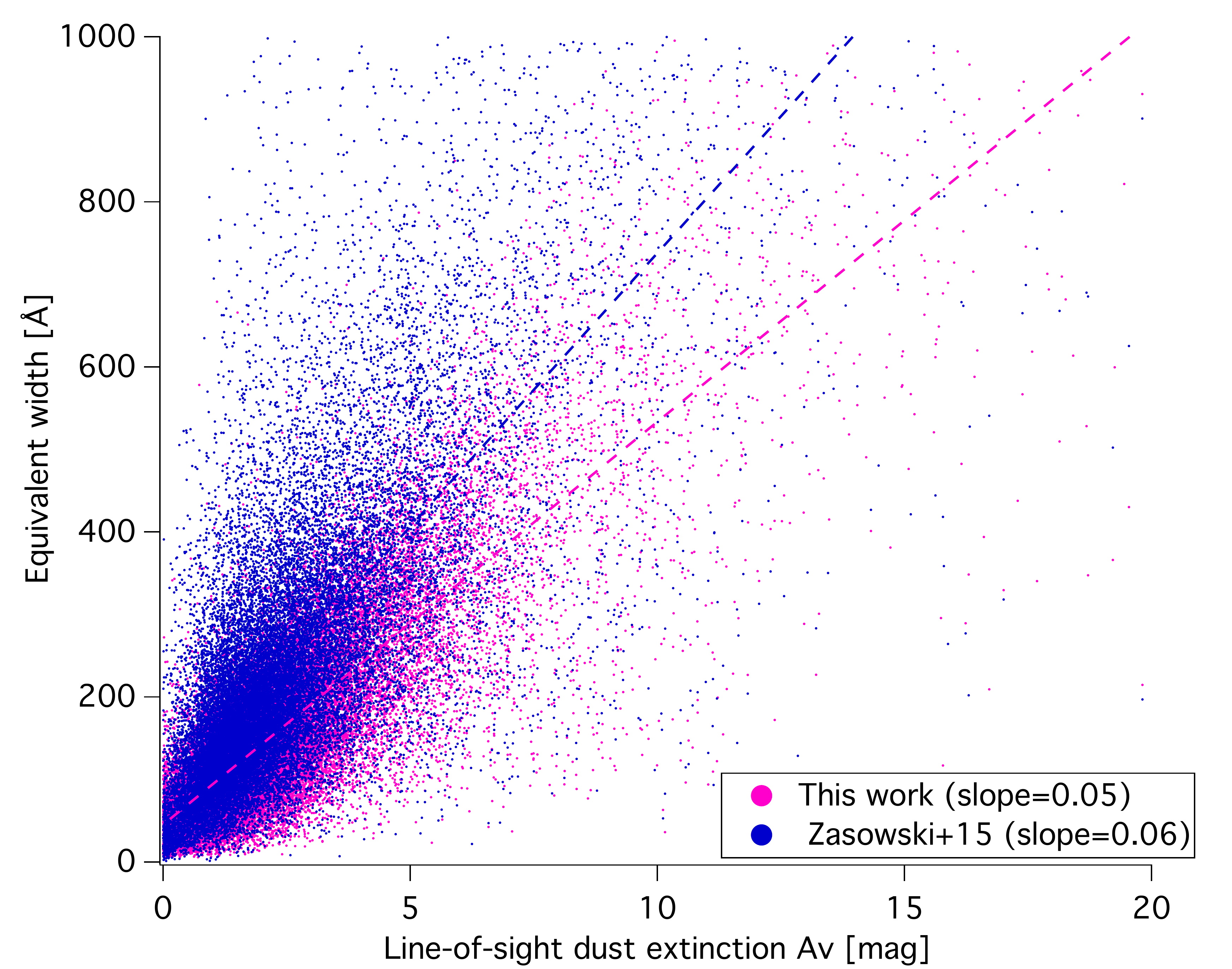}
    \caption{Comparison between the DIB EWs derived in this work and those from the \cite{Zasowski15} catalog. Both values are plotted as a function of the Cannon extinction value A$_{V}$. Linear fits to both datasets (dashed blue and pink lines) are superimposed.}
    \label{fig:myew_z15ew_vs_cannon}
\end{figure}

\begin{figure}
   \centering
    \includegraphics[width=0.45\textwidth]{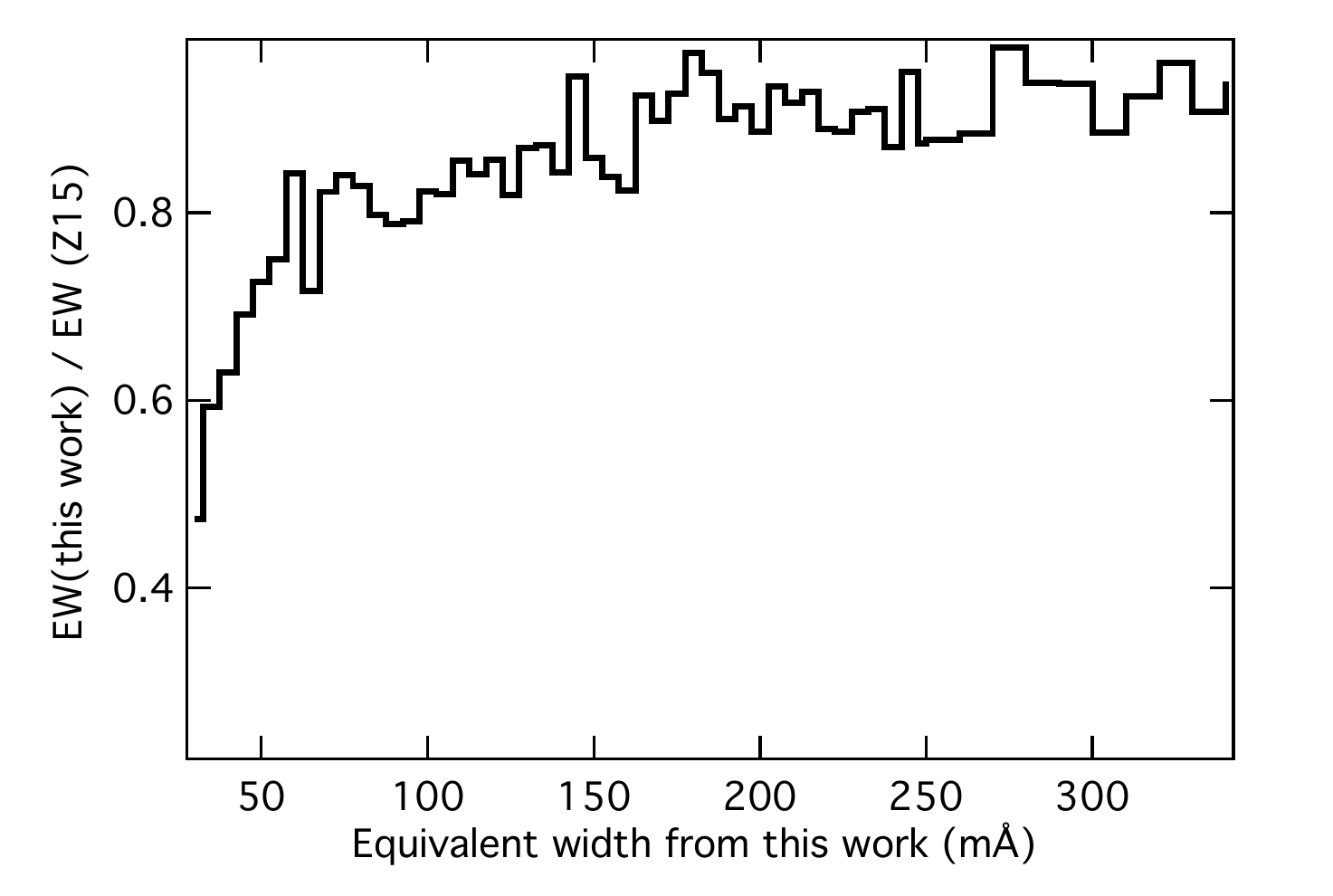}
    \caption{Average ratio between the EW derived in this work using the Cannon models and the EW derived by \cite{Zasowski15} based on ASPCAP models, here in  bins of EWs from our work. The ratio is increasing and reaches an asymptotic value on the order of 0.95 for large EWs above 300m$\mbox{\AA}$.}
    \label{fig:bin_EWs}
\end{figure}

\begin{figure}
   \centering
    \includegraphics[width=0.45\textwidth]{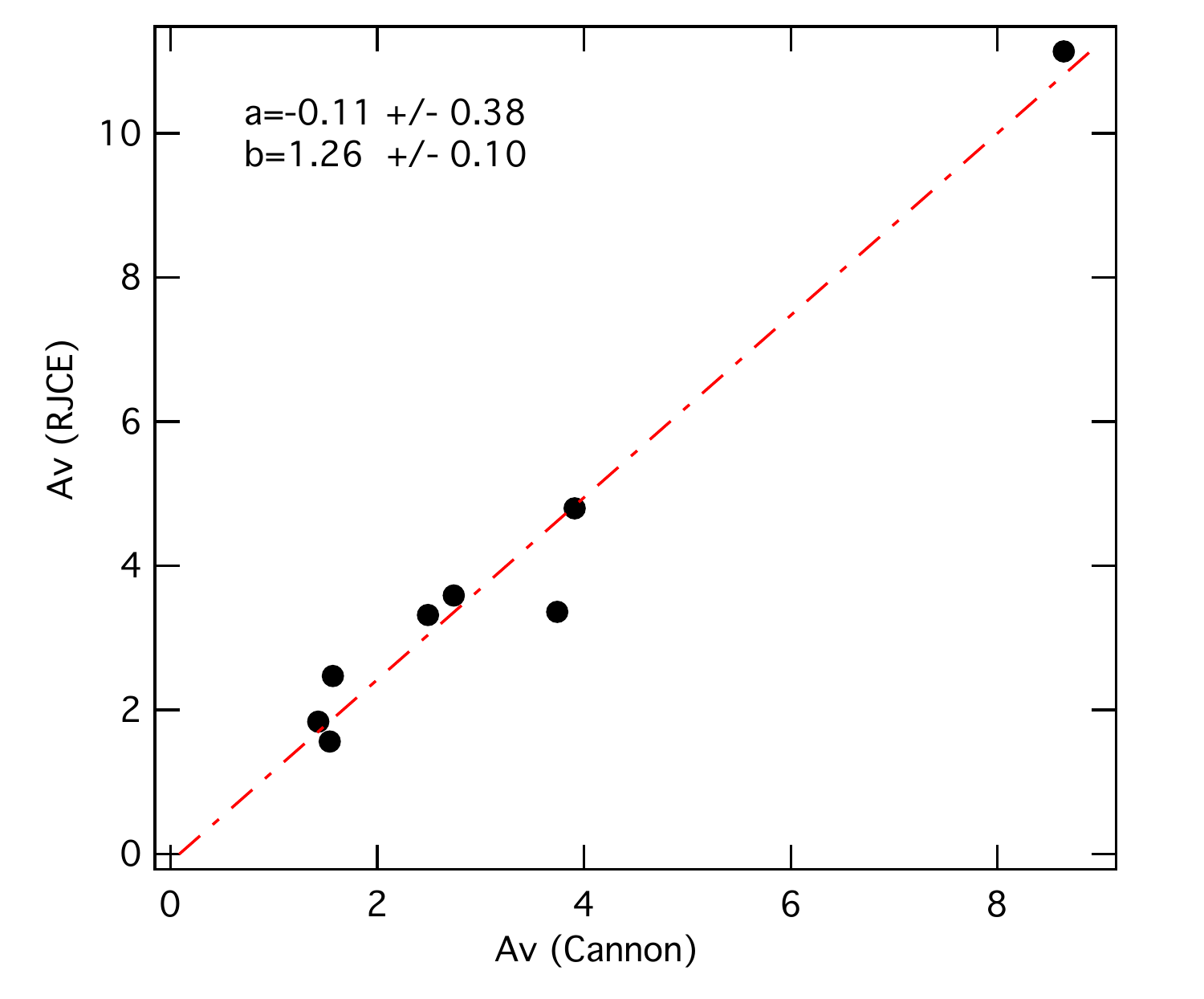}
    \caption{Cannon and RJCE extinction estimates for target stars from Table 1 from \cite{Zasowski15}. The RJCE estimate used by these authors and listed in their table is $\simeq$ 25\% higher than the Cannon estimate used in the present work.}
    \label{fig:RJCE_CANNON}
\end{figure}
\subsection{Automated fitting method}
We used the APOGEE DR 14 normalized spectra and their associated best fit \emph{Cannon} models  to extract the 15\,273 $\mbox{\AA}$\: DIB through simple Gaussian profile-fitting. The general principles are the same as in the case of the automated code applied to TSSs of APOGEE DR12 and are described in \citetalias{Elyajouri16}. Here we kept the same fitting technique and the same method of classification as in \citetalias{Elyajouri16} and \citet{Capitanio2017}, \emph{i.e.} we fitted the data to the product of an adjusted synthetic spectrum $S_\lambda^\alpha$ (here \emph{Cannon} instead of ASPCAP), a Gaussian DIB model $DIB[\sigma,\lambda,D]$ and a smooth continuum $(1+[A] \times \lambda)$ as shown in Eq.~\ref{eq:fit1}.
\begin{equation}
M_\lambda= [S_\lambda]^\alpha \times DIB[\sigma,\lambda,D] \times (1+[A] \times \lambda)
\label{eq:fit1}
\end{equation}
where each DIB is modeled as a Gaussian
function with three free parameters associated to its Gaussian RMS width ($\sigma$), central wavelength ($\lambda_{c}$) and depth ($D$) as follows:
\begin{equation}
DIB[\sigma,\lambda,D] = 1 + D \times e^{\frac{-(\lambda-\lambda_c)^2}{2\sigma^2}}
\label{eq:fit2}
\end{equation}

Prior to the final global fit, the DIB extraction code performs preliminary series of automated tests that allow to measure the residuals between the data and the adjusted stellar model in three spectral regions: the local region around the DIB A = [$\lambda_{c}$ - 10--$\lambda_{c}$ + 10] $\mbox{\AA}$, a region free of DIBs and telluric residuals B = [15\,911--15\,966] $\mbox{\AA}$\: and a narrower region free of DIBs and telluric residuals and closer to the DIB  C = [15\,200--15\,239] $\mbox{\AA}$. The associated standard deviations R$_\mathrm{A}$, R$_\mathrm{B}$ and R$_\mathrm{C}$ are used for selection criteria and estimates of the DIB uncertainties. During the fit itself the adjusted Gaussian curve representing the DIB is rejected or selected  depending on its depth $D$, its width $\sigma$ and spectrum quality criteria derived from the preliminary tests. 

\begin{itemize}
\item [$\bullet$] Stellar model adjustment $S_\lambda^\alpha$: the \emph{Cannon} stellar models are optimized for red giant stars, and they perfectly reproduce the observed data. It was not the case of the TSSs that are bluer stars for which it was necessary to adjust the stellar models.  Here we simplified our constraints on the adjustment of the stellar synthetic spectra. We still allow for a global increase or decrease of all stellar lines, however the prior value for the adjustment coefficient $\alpha$ is always unity. The fitted coefficient is always very close to one and the continuum is also always very close to a straight line of value 1.

\item [$\bullet$] DIB depth upper limit: TSSs are nearby stars whereas the APOGEE main targets can be very distant and DIBs can be much stronger. For this reason no upper limit was applied to the DIB depth.

\item [$\bullet$] DIB depth lower limit: the stellar model is so close to the data in all spectral regions that the standard deviation R$_\mathrm{A}$ of the data-model residuals in the region close to the DIB (region A) is a good enough estimate of the measurement error on the DIB depth. As a consequence the new acceptance criterion is no longer based on a comparison between the DIB depth and the standard deviations of residuals in regions A, B, and C, but is now solely: Depth $D$ $\geq$ R$_\mathrm{A}$.

\item [$\bullet$]  DIB radial velocity range: TSSs being nearby stars, the absorbing interstellar matter for those targets is characterized by a narrow radial velocity range. This is no longer the case here for the distant giant stars. As a consequence we widened the allowed interval on radial velocities with respect to the one used for TSSs. Specifically, we now restricted the spectral range for the DIB center to [15\,260--15\,286] $\mbox{\AA}$.  

\item [$\bullet$]  DIB width upper limit: we no longer used HI 21cm data to limit the total width of the DIB, instead we used limits based on the FWHM histograms from our study based on TSSs and \citet{Zasowski15}. The latter correspond to a broad range of star distances, broad enough to be valid here.

\item [$\bullet$] DIB width lower limit: we kept the same lower limit of 0.7 $\mbox{\AA}$ on the DIB width, supposed to be close to the intrinsic width as determined by \cite{Zasowski15} and \citetalias{Elyajouri16} and with some allowance for noise-related uncertainty. 
\end{itemize}
Fig.~\ref{fig:my_dr14} shows some examples of our fitting method.

\subsection{DIB upper limit estimate in case of no convergence or no DIB detection}
If the adjustment described previously failed by absence of convergence (we call it the \emph{non-convergence} case), or there was convergence but the fitted absorption is found to be smaller than the noise level (we call it the  \emph{upper-limit} case, although, as we will see, for some spectra, at the final stage of analysis, this term is not appropriate), then we re-initiated a second fitting procedure with the goal of extracting equivalent widths or upper limits in a more efficient way with more appropriate initial conditions. The total number of targets falling in the two categories and re-analyzed is 45\,393.  
First, we fitted the combination of a continuum and a depth-variable stellar model to the data (i.e, removing a DIB contribution from the global model), this time masking the DIB area. In case of convergence, this ensures that the fitted continuum and the synthetic spectrum reproduce well the data outside the DIB.
\begin{itemize}
\item[$\bullet$] \emph{Case 1}: If the fit fails again -- this is the case when the model is not adapted to the data or if there is a problem of continuum, a new standard deviation of residuals is calculated based on the difference between the data and the initial model provided by the APOGEE pipeline. 
\item[$\bullet$] \emph{Case 2}: If the fit converges then the new residuals (difference between the data and the no-DIB adjusted model) are used for an updated estimate of the standard deviation.
\end{itemize}
The second step consists in fitting a simple Gaussian to the data-model residuals in the region of the DIB. To do so, we use the same initial guesses and constraints used for the automated fitting method as described in \citetalias{Elyajouri16}.

\begin{itemize}
\item[$\bullet$] If the Gaussian fit does not converge, either because it has not found a solution or because there is a real absence of detectable DIB, then a linear fit is performed \emph{i.e,} on the continuum over the wavelength range chosen for the DIB EW [15\,260-15\,286] $\mbox{\AA}$. The selected value for the EW is then  zero and an upper limit on EW is estimated from the standard deviation of the residuals around the linear fit  Std$_{dev}$ and  the  average value of the width  of the 15\,273 $\mbox{\AA}$ DIB taken as 1.7 $\mbox{\AA}$:

\begin{equation}
EW= 0 \pm 1.7 \times \sqrt{2 \pi} \times Std_{dev}  
\label{eq:upperlinear}
\end{equation}

Some representative examples are shown in Fig.~\ref{fig:upperlimit_caselin}.
\item[$\bullet$] If the fit finds a solution, then the coefficients  of the fitted Gaussian ($\sigma_{gauss}$; D$_{gauss}$) will be used either as DIB parameters in case 1 or as initial values for a new automated fit as detailed in \citetalias{Elyajouri16} in case 2. Equivalent widths in both cases will be derived as  
EW= $\sqrt{2 \pi} \: D\:  \sigma$ where the widths are $\sigma=\{\sigma_{gauss};\sigma_{fit}\}$ and the depths are D=\{D$_{gauss}$;D$_{fit}$\} respectively. 
The errors are extracted as follows: 
\begin{equation}
Error_{DIB}=2\sqrt{\pi} \times\sigma\times Error_{depth} + 2\sqrt{\pi}\times \sigma \times Std_{dev}
\label{eq:erros_upper}
\end{equation}
\end{itemize}
These cases are illustrated in Fig.~\ref{fig:upperlimit_case2}.
Finally, for better understanding of our algorithm, Fig.~\ref{fig:flowchart_upper} shows a graphical representation that maps out our step-by-step process to determine the upper limits. It is interesting to note that this additional treatment is quite productive: for $\sim$ 2/3 of cases, the change of initial guesses for the DIB in the fitting procedure allows the Levenberg-Marquardt algorithm to converge to a good solution and a $\chi^{2}$ significantly smaller than for the first fit. This is explained  by the existence of large Doppler shifts between the star and the absorbing ISM. As a matter of fact, in our first adjustment the initial condition for the DIB center was a null radial velocity of the absorbing medium in the stellar frame. In case of large differences between the star radial velocity and the absorbing ISM radial velocity, the initial position of the DIB center was accordingly distant from the actual DIB center, and in some cases the adjustment could subsequently stop at a secondary $\chi^{2}$ minimum. This no longer happens in the second adjustment with an initial guess for the DIB center now very close to the best fit at the primary $\chi^{2}$ minimum. Similarly, peculiar or poorly modeled stellar features could favor a secondary $\chi^{2}$ minimum during the first adjustment, when they were combined with an initial position of the DIB far from the actual one. Using instead an initial guess much closer to the solution leads to the optimal solution.

\subsection{The catalog}
The extraction procedure allowed to extract EW measurements of the 15\,273 $\mbox{\AA}$\: DIB for 124\,064 different sight-lines. The catalog of equivalent widths and associated uncertainties, radial velocities and widths will be made available at the Strasbourg Data Center (CDS) after merging with the updated catalog of DIBs in the TSS spectra of Data Release 14, when available.  

Note that, because the DIB is a weak absorption, despite the high quality of the spectra and stellar models, measurement uncertainties on the EW become on the order of the EW itself for EW $\lesssim$ 0.02-0.05 $\mbox{\AA}$. This corresponds to A$_{V}\lesssim$ 0.5-0.8 mag according to the average relationship (see below).  This will result in some limitations in our analyses and correlative studies.

\section{DIB and dust extinction} \label{correl_ext}
Taking advantage of evolutions and improvements of models and spectra, we have revisited the link between 15\,273 $\mbox{\AA}$\: DIB strength and the extinction. 

To do so, we cross-matched our catalog of 15\,273 $\mbox{\AA}$\ DIBs (section \ref{extraction}) and the Starhorse \emph{Cannon} database using \texttt{TOPCAT} \citep{Taylor2005}. We removed stars that are flagged with \texttt{NUMMODELS\_BAD}, \texttt{EXTINCTION\_BAD} and \texttt{BRIGHT2MASS\_WARN} in Starhorse catalog. We finally obtained a number of 124\,064 stars with DIB detection and photometric extinction and distance.

Figure \ref{fig:ew_av} shows the equivalent width of the DIB as a function of the extinction along the sightline to the target. As expected and firstly discovered by \citet{Zasowski15} for this DIB, there is a strong correlation between these two quantities that extends over more than two orders of magnitude. We computed the median value of the EW in extinction bins of 1.6 magnitudes, and fitted this median value to a power-law relationship with the extinction bin centers. Using the new catalog we find the following relationship: 

\begin{equation}
EW_\mathrm{DIB}=  0.056 \times A_\mathrm{V}^{0.99 \pm 0.03}\: \pm 0.001\:(\mbox{\AA})
\label{eq4}
\end{equation}
for the range of A$_\mathrm{V}$=[0.55-13.35] mag. As explained above, our lower threshold is associated with EW error intervals. On the other hand, the upper limit for the linear fit is constrained by the limited number of targets with high extinctions and chosen to avoid poor statistics.
This relationship is significantly different from the previous power-law formula derived by \citet{Zasowski15} found over three orders of magnitude in both Av and EW: 
\begin{equation}
EW_\mathrm{DIB}=  0.102 \times A_\mathrm{V}^{1.01}\:(\mbox{\AA})
\label{eq:zasowski}
\end{equation}

\begin{figure*}
    \centering
   \includegraphics[width=\textwidth,height=0.68\textwidth]{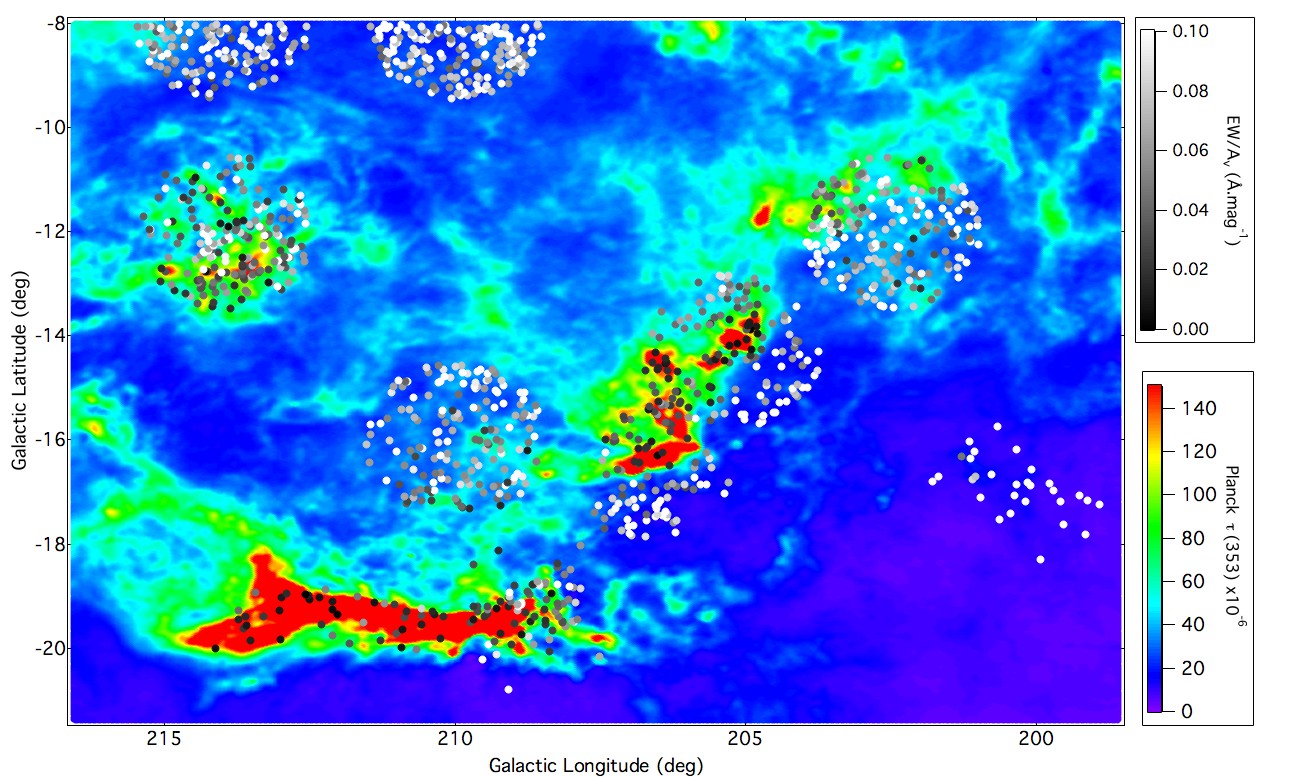}
   \caption{2D map of the Galactic interstellar dust in the Orion region (Planck 353 
GHz optical depth, color scale at bottom right). Superimposed are the SDSS/APOGEE target stars contained in the field of view. The black and white coding of the markers follows the ratio between the 15\,273 $\mbox{\AA}$ DIB strength 
and the extinction of the target light in the visible, with white (resp. black) indicative of a high (resp. low) ratio. Towards the cores of the 
clouds (in red) there is a spectacular decrease of the DIB/extinction ratio. At the cloud periphery, and also outside the clouds that are visible in the Planck image, the ratio is enhanced, reflecting the so-called "skin effect".}
    \label{fig:my_planck}
\end{figure*}
\begin{figure*}
    \centering
   \includegraphics[width=\textwidth,height=0.52\textwidth]{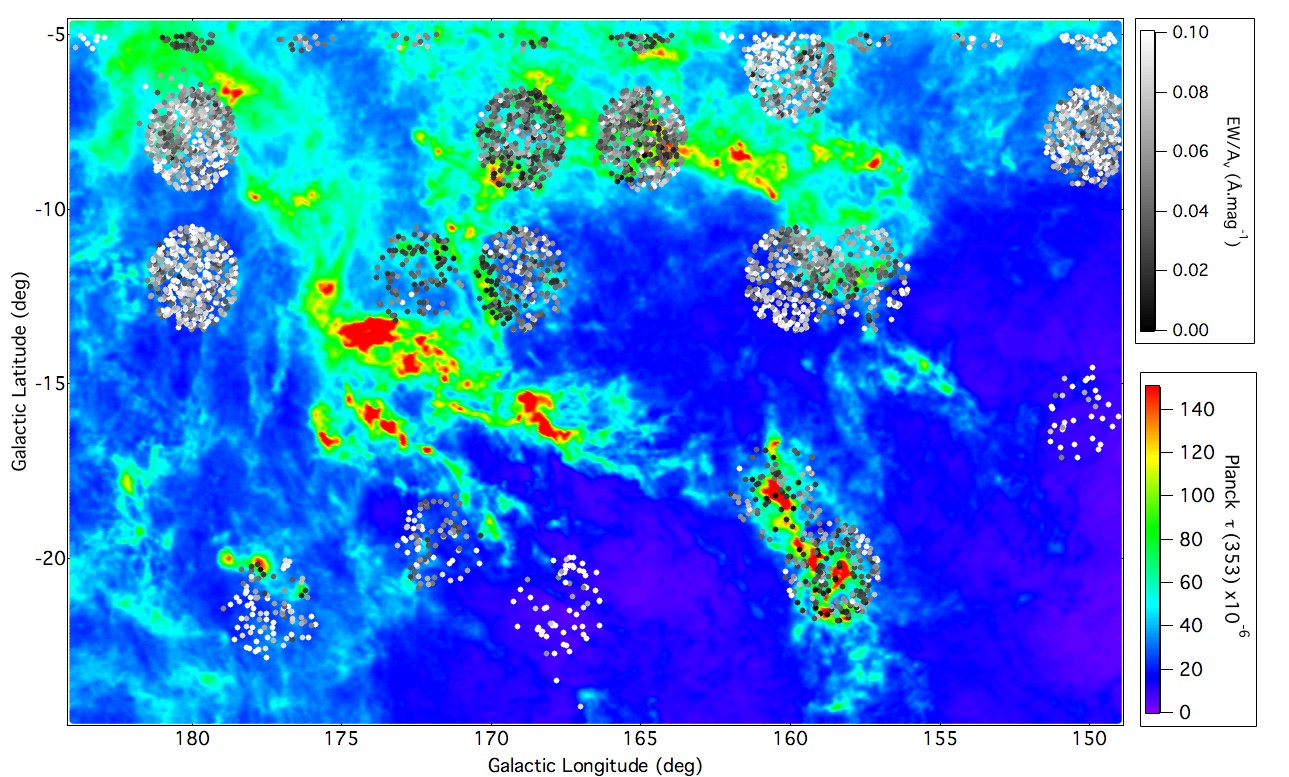}
   \caption{Same as \ref{fig:my_planck} for the Taurus region.}
    \label{fig:my_taurus}
\end{figure*}
\begin{figure*}
    \centering
   \includegraphics[width=\textwidth,height=0.48\textwidth]{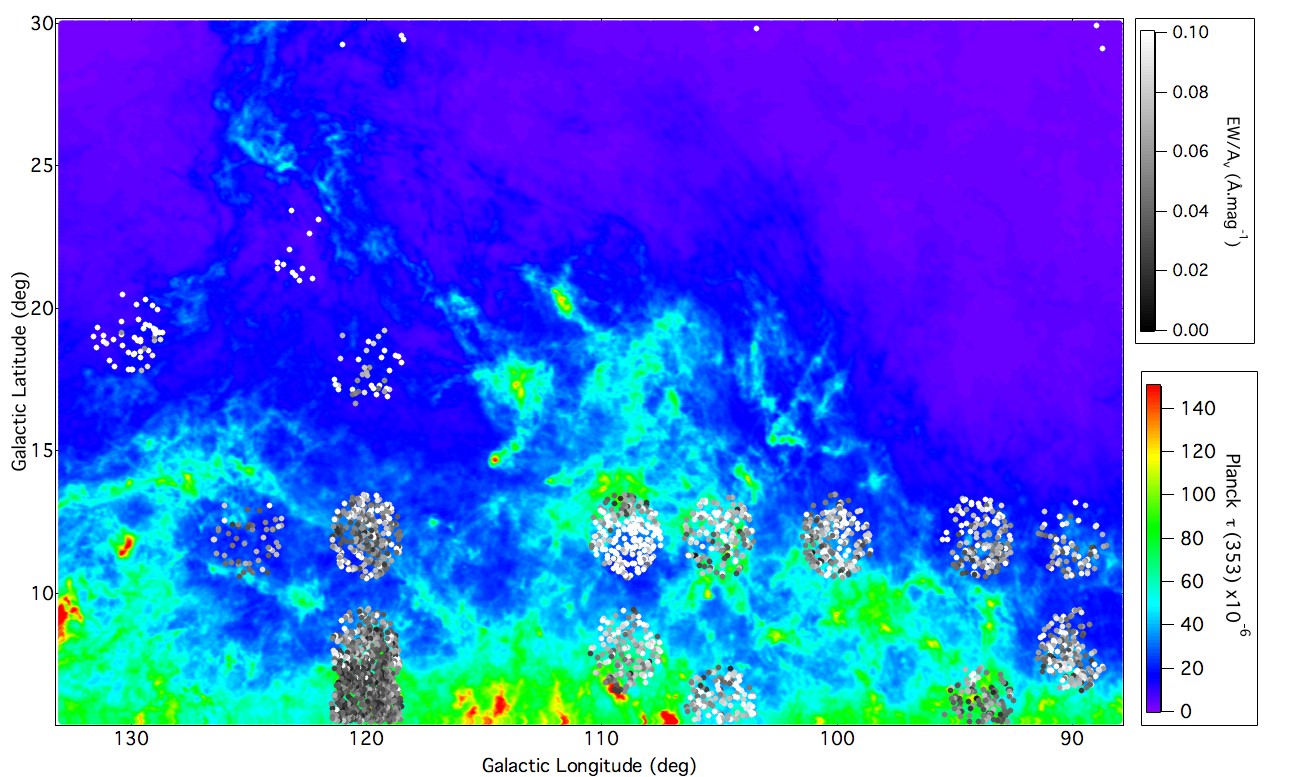}
   \caption{Same as \ref{fig:my_planck} for the Cepheus region.}
    \label{fig:my_cepheus}
\end{figure*}

We have investigated the reasons for this difference by comparing the EW results for the same individual stars and by visual inspection of the corresponding spectra and models. Fig.~\ref{fig:my_aspacap_cannon} shows  such an individual comparison. The figure shows that there are changes in the adjusted model spectra in many regions, and that by chance one of the most significant differences happens at the location of the 15\,273 $\mbox{\AA}$\: DIB. The \emph{Cannon} stellar lines are deeper than the corresponding ASPCAP lines,  and as a result the fitted DIB has a smaller EW. We further investigated the impact of these differences by cross-matching the catalog of \cite{Zasowski15} and our catalog for the same target stars. Fig \ref{fig:myew_z15ew_vs_cannon} displays the EWs derived by \cite{Zasowski15} and in this work based on the ASPCAP and Cannon stellar spectra respectively. Both are displayed as a function of the same extinction estimate, here the Cannon value. The slope of the average linear relationship between EW and the extinction is found about 15\% smaller in our case compared to the \cite{Zasowski15} derivations. This may be due to the deeper stellar lines in the Cannon spectral model and the subsequent weaker EWs. This interpretation is strongly reinforced by the observed dependence of the discrepancy between the two determinations on the DIB strength, shown in Fig \ref{fig:bin_EWs}. The figure displays the average of the ratio between our determination and the one of \cite{Zasowski15} in bins of increasing EWs. It is clearly seen that, the stronger the DIB EW, the smaller the difference between the two determinations. We subsequently conclude that a non negligible fraction of the difference between our two relationships can be attributed to the use of different synthetic stellar models.

A second reason for the discrepant slopes is the use of different extinction estimates. Fig. \ref{fig:RJCE_CANNON} displays the RJCE and Cannon values of the extinction A$_{V}$ for the target stars listed in the table 1 of \cite{Zasowski15}. The average linear relationship between the two estimates is shown in the figure. According to it, RJCE values are on the order of 25\% above the Cannon estimates for the set of stars.  The two effects, the one due to the use of the Cannon extinction instead of the RJCE estimate on the one hand, and the one due to the slightly different  Cannon spectrum on the other hand, both result in a global decrease of the EW to extinction ratio and explain why our DIB to extinction ratio is about 40\% below the one found by \cite{Zasowski15}.

Fig.~\ref{fig:ew_av} also shows that the linear relationship is no longer valid at high A$_{V}$ values, and that an increasing number of DIBs have EWs that are significantly lower than predicted by this law. This is the so-called \emph{skin effect} first detected by \citet{Snow1974} and then often found in the spectra of highly reddened stars. This effect is here illustrated in Fig. \ref{fig:ew_av} using averages of DIB EWs in bins of 1.6 mag of A$_{V}$ extinction. This choice of A$_{V}$ intervals is a compromise between the need for a minimum number of targets per bin and the maximum discretization over the whole range. 

This depletion  is better seen  when zooming at the break region for high A$_{V}$ values (right part of Fig.~\ref{fig:ew_av}). 
The ratio between the median value of column of DIB carrier and the extinction (column of grains) is decreasing while A$_{V}$ is increasing. This is clearly due to an increasing number of sightlines with relatively weak DIBs. On the other hand, a large fraction of the sightlines have strong DIBs that remain quasi-proportional to the extinction.

\begin{figure*}
    \centering
        \includegraphics[width=\textwidth]{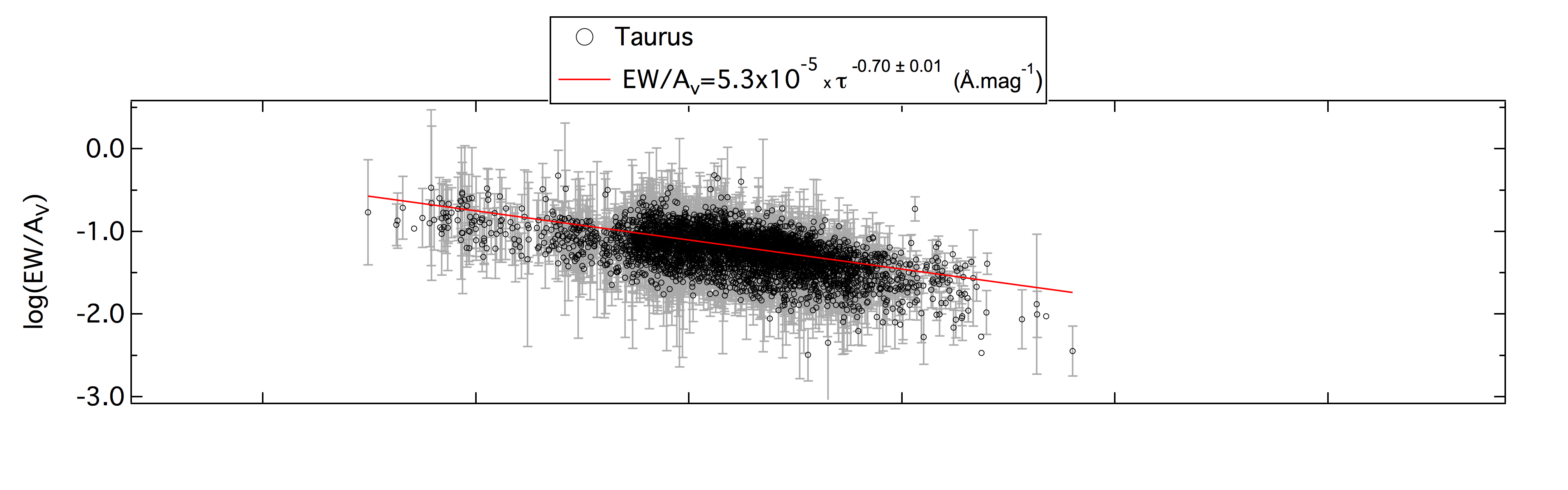}
        \includegraphics[width=\textwidth]{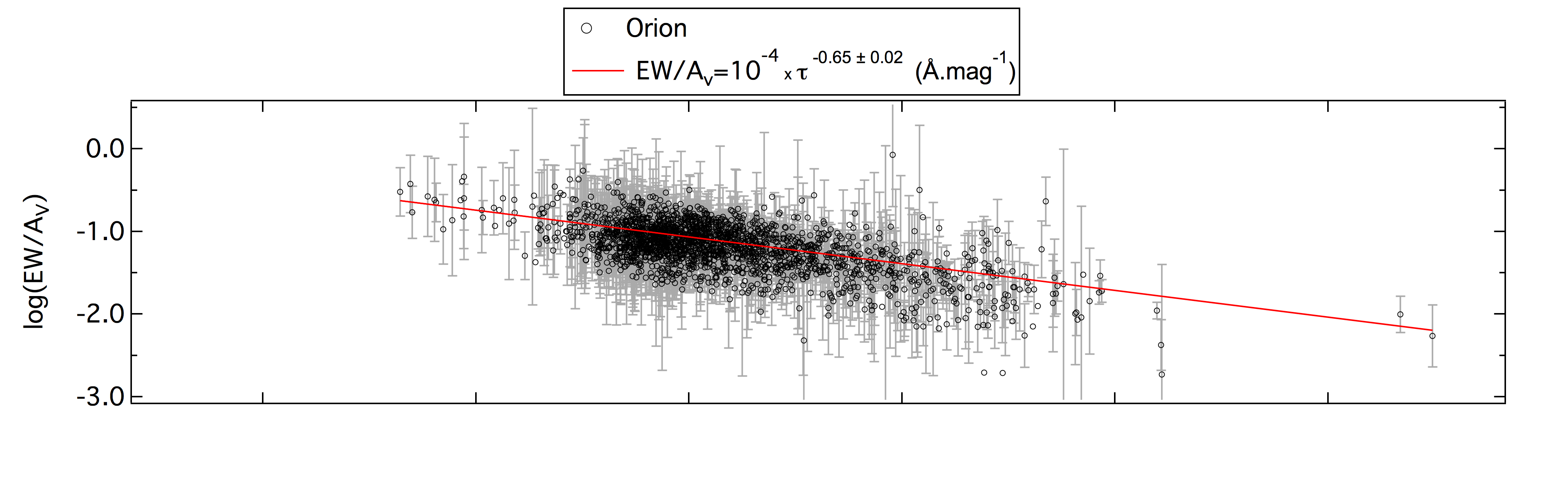}
        \includegraphics[width=\textwidth]{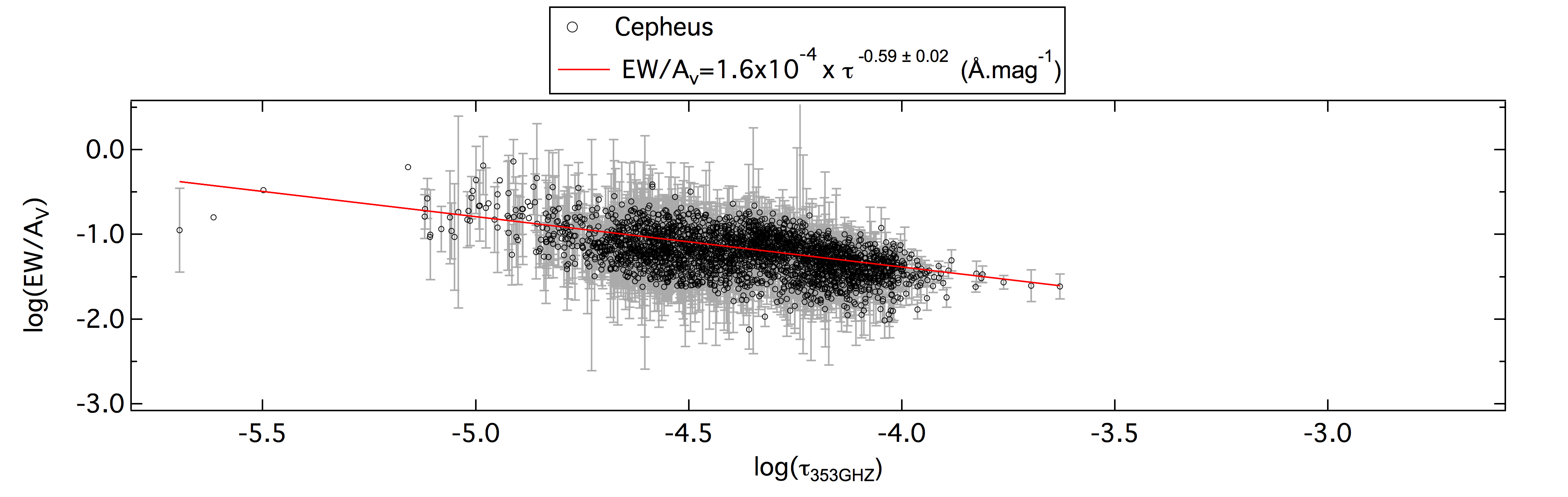}
    \caption{DIB Equivalent width to extinction A$_{V}$ ratio  towards stars behind Taurus (top), Orion (middle) and Cepheus (bottom) cloud complexes, as a function of the dust optical thickness measured by Planck $\tau$(353 GHz). Linear fits in the Log-Log scale are shown in red.}
    \label{fig:rapdibext_horsplan}
\end{figure*}

\begin{figure*}
    \centering
        \includegraphics[width=0.8\textwidth]{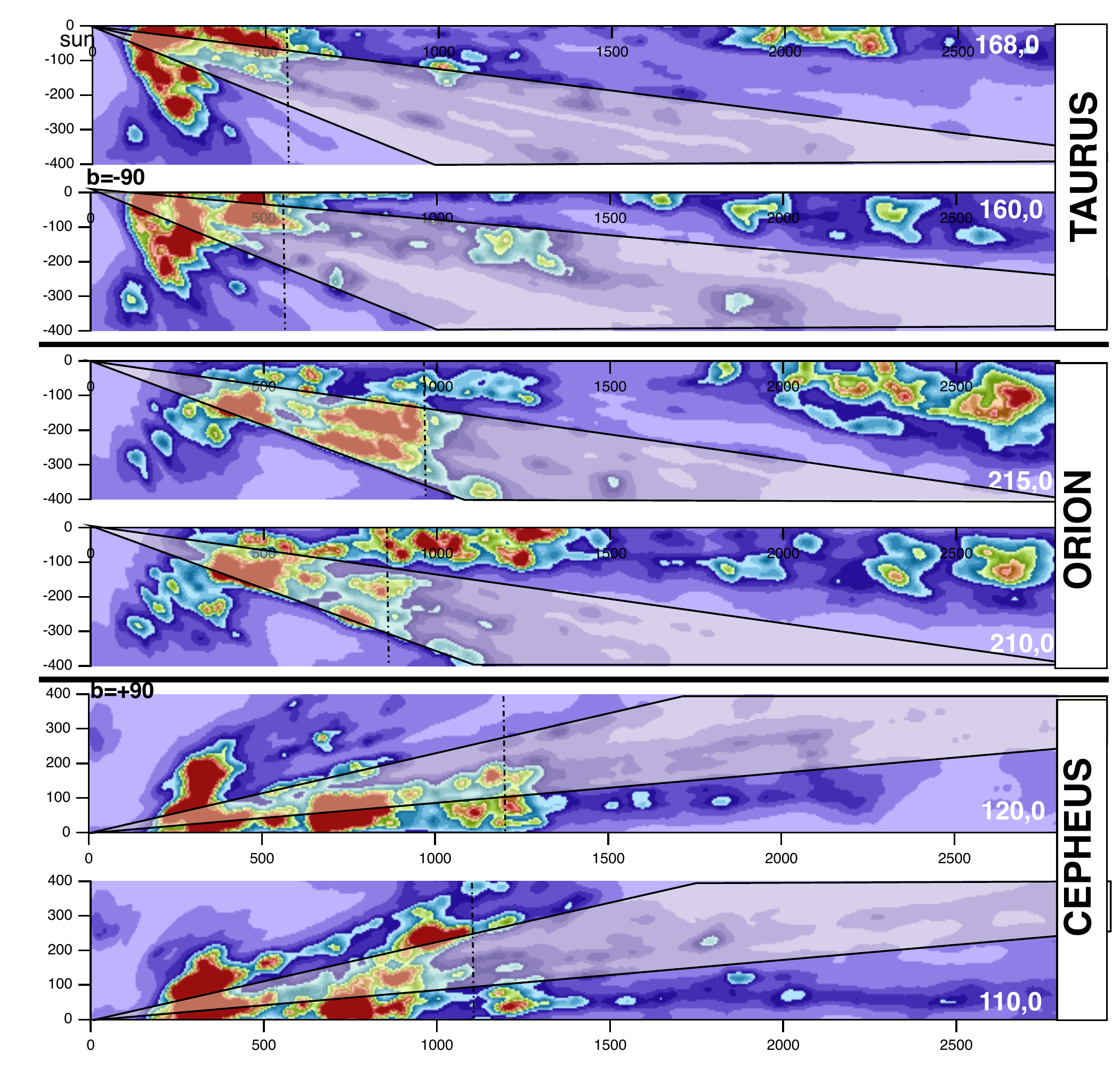}
    \caption{3D distribution of dust in vertical planes containing the Sun and crossing the Taurus, Orion and Cepheus clouds. The North Galactic pole is to the top in each figure. The cloud distribution is based  on \cite{Lallement2019}. For each cloud group two vertical planes are selected in such a way the number of target sightlines falling very close to at least one of the two planes is maximized. The Galactic longitudes corresponding to each plane are indicate at bottom right of each figure. The dashed black line encloses the densest structures (see text).}
    \label{fig:cut3D}
\end{figure*}

It is interesting to compare the skin effect seen here in a large variety of sightlines with the specific case of a dark cloud, \emph{i.e.}  a quasi-perfect mono-cloud situation in which all the extinction is generated in a unique dense cloud. To do so, we have added to the figure the results obtained in \citet{Elyajouri2017b} for the Barnard 68 dark cloud. The corresponding DIB EW data points correspond to some of the largest departures from the values predicted by the linear relation. However, they are still comparable to tens of other similar cases, suggesting that the marked disappearance of DIBs in Barnard 68 is far from an exception. 

\section{The DIB carrier depletion illustrated with Planck} \label{skin_effect}

A fraction of the APOGEE sightlines is out of the Plane and crosses the nearby clouds of the Gould belt. For those targets that are located behind the main clouds, the observed dust emission is entirely or almost entirely generated by the nearby clouds. For this reason it is interesting to investigate the link between the DIB to extinction ratio and the dust emission. One of the goals is to detect variations among the cloud complexes or conversely some general laws for the skin effect.

Fig.~\ref{fig:my_planck} is a map of the Planck optical depth $\tau_{353}$ in the region of the nearby Orion clouds (data from \citealt{Planck2016}). Superimposed on the map are the APOGEE targets, with markers color-coded according to their EW/A$_\mathrm{V}$ ratio. To ensure that the stars are located behind the Orion clouds, targets closer than 600 pc were eliminated. In addition, we discarded very low reddening targets and/or very low DIB equivalent widths  in order to avoid the corresponding uncertain ratios. The corresponding thresholds are  A$_\mathrm{V}$ $\geq$ 0.2 mag and EW $\geq$ 0.03 $\mbox{\AA}$. The EW DIB decrease in the dense phase is conspicuously illustrated in this figure: towards all cloud cores ($\tau_{353}$ $\geq$ 10$^{-5}$) the DIB to extinction ratio decreases dramatically. Reciprocally, in the more diffuse regions the ratio reaches values that are above the average ratio of $\sim$0.06 $\mbox{\AA}$ mag$^{-1}$. This is the first time such a clear visualization is possible, and it is allowed by the combination of the high-quality APOGEE data and the high quality and spatial resolution of the Planck maps. Similar figures, this time for the Taurus and Cepheus clouds are presented in Fig.~\ref{fig:my_taurus} and Fig.~\ref{fig:my_cepheus}. The same restrictions were applied on A$_\mathrm{V}$ and EW, and the lower limits for the target distances were chosen as 1 kpc for Taurus and 1.3 kpc for Cepheus. The same striking association between cloud cores and low DIB-extinction ratios is immediately visible. 

A visual inspection of the three figures does not reveal any marked difference between the three cloud associations. In order to quantify better the DIB level off we have represented in Fig.~\ref{fig:rapdibext_horsplan} log(EW/A$_\mathrm{V}$)  as a function of log($\tau_{353}$) for the three regions and the same selected targets located beyond the clouds. A first attempt to quantify the level off was made by
means of a linear fit in the log-log scale for the merged datasets. The resulting best fit relationship between EW/A$_\mathrm{V}$ and $\tau_{353}$ was found to be:
\begin{equation}
EW/A_{V}=7.6\times 10^{-5} \times \tau_{353}^{-0.66\pm0.01}\:(\mbox{\AA} \mathrm{mag}^{-1})
\label{eq:skineffect}
\end{equation}

We also performed the same adjustments for the three regions separately, this time taking into account uncertainties on the DIB to EW ratio (the fractional error on the ratio is estimated as the square root of the quadratic sum of fractional errors on the two terms). The adjustments are displayed in Fig \ref{fig:rapdibext_horsplan} and the results were:
\begin{align}
\label{eq:skineffect_clouds}
Taurus: EW/A_{V}=5.3 \times 10^{-5} \times \tau_{353}^{-0.70\pm0.01}\:    (\mbox{\AA} \mathrm{mag}^{-1})\\
Orion: EW/A_{V}=1.0\times 10^{-4} \times \tau_{353}^{-0.65\pm0.02}\:(\mbox{\AA} \mathrm{mag}^{-1})\\
Cepheus: EW/A_{V}=1.6\times 10^{-4} \times \tau_{353}^{-0.59\pm0.02}\: (\mbox{\AA} \mathrm{mag}^{-1})
\end{align}

\begin{figure*}
    \centering
     \includegraphics[width=0.8\textwidth,height=0.75\textwidth]{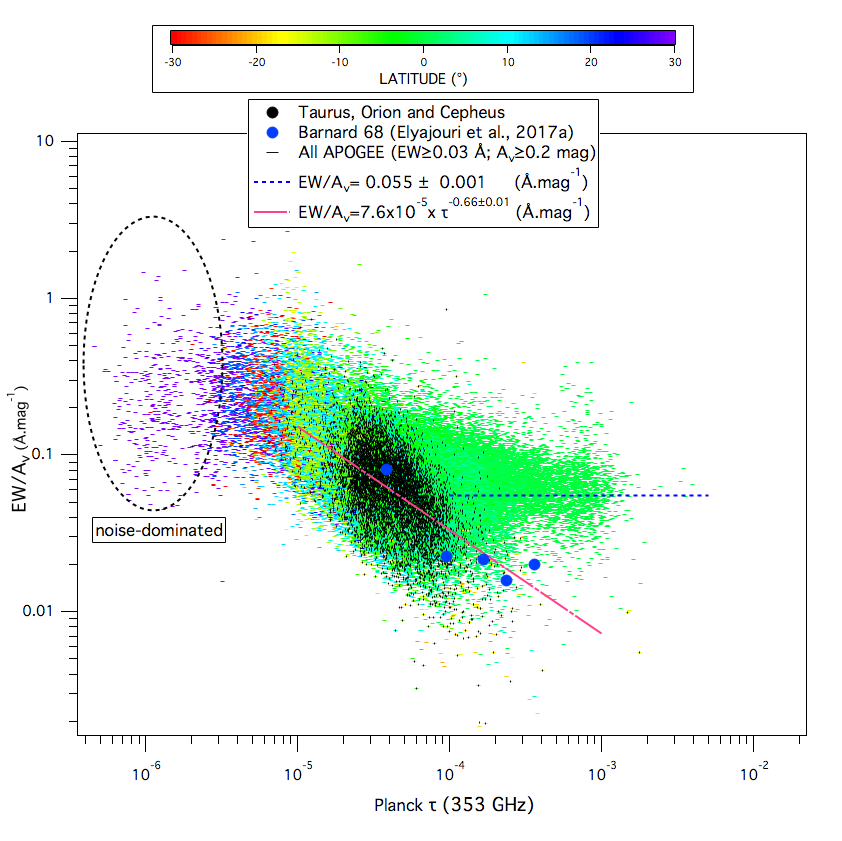}
    \caption{The DIB to extinction ratio $EW/A_\mathrm{v}$ as a function of $\tau_{353}$ for all APOGEE sightlines with EW $\geq$ 0.03 $\mbox{\AA}$\: and Av $\geq$ 0.2 mag. These limitations ensure a reliable determination of the EW/Av ratio. Orion, Taurus and Cepheus targets are  marked  by  black  signs,  while  all other targets are color-coded according to the Galactic latitude. Here, one can distinguish three different regimes: high-latitude, low dust sightlines that correspond to nearby diffuse clouds and a high $EW/A_\mathrm{v}$ of $\simeq$ 0.23 $\mbox{\AA}\:mag^{-1}$, partially mixed phases in nearby off-Plane cloud complexes following the Taurus-Orion-Cepheus trend (red curve), and distant, high dust optical depth sightlines within the midplane for which the $EW/A_\mathrm{v}$ ratio reaches a constant value of $\simeq 0.056\: \mbox{\AA}\:mag^{-1}$ (blue dashed curve).}
    \label{fig:rapdibext_all}
\end{figure*}

These quantitative results show that the average relationships present some differences from one region to the other, with a slope varying from -0.70 to -0.59 and without overlap of the allowed intervals.

In a search for the origin of such differences, we have made use of three-dimensional maps of the interstellar dust. Such maps are based on extinction measurements for stars at distributed distances and assign  locations to the main cloud complexes \citep{Lallement2019}. Fig \ref{fig:cut3D} displays the dust distribution in vertical planes crossing the clouds. For each complex, two vertical cuts are shown, for two longitudes chosen to correspond to large concentrations of target star sightlines. In each vertical plane, the range of Galactic latitudes of the target stars selected in Fig \ref{eq:skineffect_clouds} are indicated by shaded sectors. A black line delimits the maximum distance from the Sun to the densest regions in the corresponding sector. The dust differential opacity threshold has been here chosen at 1.5 mag.pc$^{-1}$. The figures show that most sightlines cross several dense structures, i.e.  a sightline does not probe a unique gas phase, instead series of dense and more diffuse regions are distributed along the sightlines. This probably explains the large dispersion of data points in Fig. \ref{fig:rapdibext_horsplan}, since DIB to extinction ratios for the various phases will be smeared out in different ways depending on the sightlines. However, interestingly, the distance from the Sun to the dense cloud limit increases from Taurus to Orion then to Cepheus (although the increase from Taurus to Orion is much stronger than from Orion to Cepheus), i.e. this distance increases while simultaneously the slope of the DIB to extinction ratio relation with the dust thickness gets shallower. An increase in the limit distances and the number of structures along the sightlines implies that the mixing of diffuse and dense regions must be stronger in Orion than in Taurus and even higher in Cepheus compared to Orion.  For those reasons, we interpret the change of slope among the three clouds shown in Fig. \ref{fig:rapdibext_horsplan} as an indicator of the strength of phase mixing. In case of Taurus, its smaller number of structures and smaller resulting phase mixing, the level off the DIB is more distinctly linked to the dust column and the slope of the decrease is stronger. This is the opposite for Cepheus. If our interpretation is correct, this implies that the actual DIB to extinction ratio decrease from the diffuse to the dense phase is even stronger than suggested by the slopes in Fig. \ref{fig:rapdibext_horsplan}.

Finally, we have attempted to use the Planck data and all APOGEE sightlines to study the DIB level off  in all types of configurations. To do so we removed all conditions on the target distances and kept only the lower limits on $A_{V}$ and EW imposed to avoid too large uncertainties on the DIB to extinction ratio.  The total number of sightlines becomes 75\,840. Fig.~\ref{fig:rapdibext_all} shows the DIB to extinction ratio as a function of $\tau_{353}$. The Orion, Taurus and Cepheus targets discussed above are marked by black signs, while all other targets are color-coded according to the Galactic latitude. Interestingly, one can distinguish three different regimes: at high latitude, thus for low dust optical depth and mostly nearby clouds, the DIB-$A_\mathrm{V}$ ratio apparently stabilizes, i.e. no longer increases with decreasing $\tau_{353}$. It reaches a limit on the order of $\simeq 0.23\mbox{\AA}\:mag^{-1}$. Such a value is probably representative of very diffuse regions located close to the boundary of the Local Cavity (see, e.g., maps from \citealt{Capitanio2017}). The second regime corresponds to the trend observed for the off-Plane Taurus, Orion and Cepheus regions, as already discussed, however this time it also applies to a large number of targets within the Plane. The third regime corresponds to large optical depths and sightlines within the mid-plane (as indicated by the low values of Galactic latitude). Here the ratio is no longer depending on $\tau_{353}$ and reaches a limit on the order of 0.05 $\mbox{\AA}$\ mag$^{-1}$, a value very close to the average found from the global linear fit to the data. This value must correspond to a strong mixing of diffuse and dense phases all along the sightlines.

Superimposed in Fig.~\ref{fig:rapdibext_all} are the Barnard 68 corresponding data points for the five regions delineated in \citet{Elyajouri2017b} that span extinctions A$_{V}$ between  2 and 20 mag. To convert A$_{V}$ into 
$\tau_{353}$ we used the average relationship A$_{V}$ $\simeq$ 4.5 10$^{4}$ $\tau_{353}$, based on the classical A$_{V}$ to E(B-V) ratio of 3.1 and the average E(B-V) vs $\tau_{353}$ relationship of \cite{Remy2018}. It can be seen from the Figure that these data points fall in the middle of the ensemble of data points for Taurus, Orion and Cepheus, \emph{i.e.,} again, the dark globule does not correspond to an exceptional situation in terms of DIB carrier depletion. This is interesting since Barnard 68 is immersed in a large opaque region with only a few ionizing stars, in strong opposition to Orion. 

\section{Summary}\label{conclusion}
We have analyzed red giants NIR spectra from the last  SDSS/APOGEE survey data release (DR14) with the goal of extracting the 15\,273 $\mbox{\AA}$ DIB. To do so, we used the code developed by \citetalias{Elyajouri16} with some adaptations motivated by the change of target types, namely from early- or intermediate-type stars to red giants. The use of the stellar models computed by means of the data-driven \emph{Cannon} algorithm greatly improves the analysis.  In the region of the 15\,273 $\mbox{\AA}$ DIB the adjustment has proven to be good enough to allow the extraction of the DIB equivalent width, width and Doppler velocity through Gaussian fitting of the interstellar absorption. Careful and severe examinations of the DIB parameters, the continuum shape and the quality of the adjustment were done. This resulted in a conservative selection of reliable DIB parameters for a total of 124\,064 lines of sight. 
We have compared our DIB measurements with their associated visual extinctions extracted from the Starhorse database. We revisited the DIB$-$A$_{V}$ relationship first reported by \citet{Zasowski15} and found the DIB strength average  to be linearly correlated with dust extinction over at least two orders of magnitude, however with a smaller equivalent width-to-extinction ratio than the one derived by these authors for the red giants and later by \citetalias{Elyajouri16} for the hotter TSS targets:
EW$_\mathrm{DIB}=  0.056 \times A_\mathrm{V}^{0.99 \pm 0.03} (\mbox{\AA})$. We showed how this variation is explained by the differences between the ASPCAP and \emph{Cannon} models and by the use of Cannon A$_{V}$ extinctions instead of A$_{V}$ extinctions deduced from the RJCE technique. 

It has also been found from the global DIB-extinction comparative study that the 15\,273 $\mbox{\AA}$\: DIB tends to weaken with respect to extinction in the spectra of stars situated behind dense regions. With the goal of going one step further in the analysis of the DIB carrier depletion in dense cores, we investigated the link between the EW/A$_{V}$ ratio and the distribution of dust as traced by Planck. In particularly, the Taurus, Orion and Cepheus cloud complexes were studied using targets distant enough to be located behind the clouds. The depletion effect is spectacularly illustrated for those cloud complexes, thanks to the high quality of the APOGEE data and of the Planck maps. We have fitted an average relationship between the DIB A$_\mathrm{V}$ ratio and the Planck optical depth $\tau_{353}$ through a log-log linear formula: $EW/A_{V}=7.6\times 10^{-5} \times \tau_{353}^{-0.66\pm0.01}\:(\mbox{\AA}\: \mathrm{mag}^{-1})$.

We have performed a separate study of the three cloud complexes in a search for potential differences in the DIB depletion effect. For the Taurus clouds (resp. Orion, Cepheus), and taking into account uncertainties on equivalent widths and extinction the coefficients of the log-log linear relationship are: $(5.3\times10^{-5};-0.70\pm0.01),(1.0\times10^{-4};-0.65\pm0.02),(1.6\times10^{-4};-0.59\pm0.02)$ respectively. Using 3D maps of the dust clouds obtained by inversion of stellar individual extinctions \citep{Lallement2019}, we have shown that the decrease of the depletion gradient from Taurus to Cepheus is very likely due to a corresponding increasing number of structures encountered along the paths to the target stars behind the three complexes, and the subsequent increase of phase mixing. Our results provide first empirical average relationships between the 15\,273$\mbox{\AA}$\: DIB carrier depletion in molecular clouds as a function of the dust optical depth.

Finally, we have used the entire dataset in an attempt to identify different depletion regimes for the APOGEE sightlines. We found that, not surprisingly, for large dust optical depths exceeding $\sim$ 10$^{-4}$, \emph{i.e.} for low latitude and extended sightlines, the DIB to A$_\mathrm{V}$ ratio tends to the mean value of about 0.05-0.06 $\times$ A$_\mathrm{V}$ that is characteristic of a strong mixing of diffuse and dense phases distributed along the extended paths. For intermediate $\tau_{353}$ values, between $\sim$ 10$^{-5}$ and 10$^{-4}$ we clearly see the same depletion effect that is affecting the nearby off-Plane cloud complexes, now extended to mi-plane targets. Finally, for low values of $\tau_{353}$ below  10$^{-5}$, the DIB to extinction ratio seems to be constant at $\sim$ 0.2 $\times$ A$_\mathrm{V}$, a value significantly above than the strong mixing average. This high value corresponds to the nearby diffuse clouds at the periphery of the local cavity. 

\section{Discussion}\label{discussion}

This break seen at high A$_{V}$ values is well known  as the skin effect \citep{Herbig95} and has been observed several times for optical DIBs. In the case of the 15\,273$\mbox{\AA}$\: DIB,  it was also observed recently in the B68 nearby dark cloud \citep{Elyajouri2017b} with a strength comparable to multiple sightlines from the present study. Very recently, it has been more specifically characterized and quantified for optical DIBs by \cite{Fan2017} who demonstrated the existence of a maximum carrier abundance with respect to dust grains for a molecular H$_{2}$ fraction on the order of 0.3.

The origin of the DIB carrier depletion in dense molecular cores is still unclear. In the context of PAH-like carriers, UV radiation would certainly directly influence the charge state of PAHs \citep[e.g.][]{Cami97,2005A&A...432..515R} and the resulting disappearance of some DIBs could simply be due to the evolution of the charge state from the outer to inner regions of the clouds.  On the other hand, photo-processing and hydrogenation or de-hydrogenation may enter in play \citep[e.g.][]{Vuong00}. More recently \citet{BerLall17} suggested that a general carrier disappearance could additionally be due to coagulation of DIB carriers into aggregates and accretion onto dust grains, as predicted by \cite{Jones2016} in the context of new models of dust. It is well known that grains grow in compact cores, generating the so-called coreshine effect \citep{Lefevre2014} and grain coalescence to form aggregates may also involve macro-molecules associated with DIBs. Future detailed studies of the DIB carrier depletion in regions with differing radiation field, pressure, temperature and grain properties will hopefully help disentangling the different contributions. In this respect, empirical relationships involving dust emission such as those established here may complement observed links with the physical state of the gas such as those of found by \cite{Fan2017} and bring additional clues.

\begin{acknowledgements}
ME acknowledges funding from the "Region Ile-de-France" through the DIM-ACAV project. RL acknowledges support from the CNRS PCMI national program.

This research has made use of the SIMBAD database, operated at CDS, Strasbourg, France.

Funding for the Sloan Digital Sky Survey IV has been provided by
the Alfred P. Sloan Foundation, the U.S. Department of Energy Office of
Science, and the Participating Institutions. SDSS-IV acknowledges
support and resources from the Center for High-Performance Computing at
the University of Utah. The SDSS web site is www.sdss.org.

SDSS-IV is managed by the Astrophysical Research Consortium for the 
Participating Institutions of the SDSS Collaboration including the 
Brazilian Participation Group, the Carnegie Institution for Science, 
Carnegie Mellon University, the Chilean Participation Group, the French Participation Group, Harvard-Smithsonian Center for Astrophysics, 
Instituto de Astrof\'isica de Canarias, The Johns Hopkins University, 
Kavli Institute for the Physics and Mathematics of the Universe (IPMU) / University of Tokyo, Lawrence Berkeley National Laboratory, 
Leibniz Institut f\"ur Astrophysik Potsdam (AIP),  
Max-Planck-Institut f\"ur Astronomie (MPIA Heidelberg), 
Max-Planck-Institut f\"ur Astrophysik (MPA Garching), 
Max-Planck-Institut f\"ur Extraterrestrische Physik (MPE), 
National Astronomical Observatory of China, New Mexico State University, New York University, University of Notre Dame, 
Observat\'ario Nacional / MCTI, The Ohio State University, 
Pennsylvania State University, Shanghai Astronomical Observatory, 
United Kingdom Participation Group,
Universidad Nacional Aut\'onoma de M\'exico, University of Arizona, 
University of Colorado Boulder, University of Oxford, University of Portsmouth, 
University of Utah, University of Virginia, University of Washington, University of Wisconsin, 
Vanderbilt University, and Yale University.  
\end{acknowledgements}
\bibliographystyle{aa}
\bibliography{mybib2.bib}

\begin{thebibliography}{53}
\expandafter\ifx\csname natexlab\endcsname\relax\def\natexlab#1{#1}\fi

\bibitem[{{Abolfathi} {et~al.}(2018){Abolfathi}, {Aguado}, {Aguilar}, {Allende
  Prieto}, {Almeida}, {Tasnim Ananna}, {Anders}, {Anderson}, {Andrews},
  {Anguiano}, \& et~al.}]{Abolfathi2018}
{Abolfathi}, B., {Aguado}, D.~S., {Aguilar}, G., {et~al.} 2018, \apjs, 235, 42

\bibitem[{{{\'A}d{\'a}mkovics} {et~al.}(2005){{\'A}d{\'a}mkovics}, {Blake}, \&
  {McCall}}]{Adamk05}
{{\'A}d{\'a}mkovics}, M., {Blake}, G.~A., \& {McCall}, B.~J. 2005, \apj, 625,
  857

\bibitem[{{Adamson} {et~al.}(1991){Adamson}, {Whittet}, \&
  {Duley}}]{Adamson1991}
{Adamson}, A.~J., {Whittet}, D.~C.~B., \& {Duley}, W.~W. 1991, \mnras, 252, 234

\bibitem[{{Anders} {et~al.}(2018){Anders}, {Queiroz}, {Chiappini}, {Santiago},
  {Fern{\'a}ndez-Trincado}, {Meza}, \& {SDSS-IV/APOGEE
  Collaboration}}]{Anders2018}
{Anders}, F., {Queiroz}, A.~B., {Chiappini}, C., {et~al.} 2018, in IAU
  Symposium, Vol. 334, Rediscovering Our Galaxy, ed. C.~{Chiappini},
  I.~{Minchev}, E.~{Starkenburg}, \& M.~{Valentini}, 153--157

\bibitem[{{Bertaux} \& {Lallement}(2017)}]{BerLall17}
{Bertaux}, J.-L. \& {Lallement}, R. 2017, \mnras, 469, S646

\bibitem[{{Cami} {et~al.}(1997){Cami}, {Sonnentrucker}, {Ehrenfreund}, \&
  {Foing}}]{Cami97}
{Cami}, J., {Sonnentrucker}, P., {Ehrenfreund}, P., \& {Foing}, B.~H. 1997,
  \aap, 326, 822

\bibitem[{{Campbell} {et~al.}(2016){Campbell}, {Holz}, {Maier}, {Gerlich},
  {Walker}, \& {Bohlender}}]{Campbell16}
{Campbell}, E.~K., {Holz}, M., {Maier}, J.~P., {et~al.} 2016, \apj, 822, 17

\bibitem[{{Capitanio} {et~al.}(2017){Capitanio}, {Lallement}, {Vergely},
  {Elyajouri}, \& {Monreal-Ibero}}]{Capitanio2017}
{Capitanio}, L., {Lallement}, R., {Vergely}, J.~L., {Elyajouri}, M., \&
  {Monreal-Ibero}, A. 2017, A\&A, 606, A65

\bibitem[{{Casey} {et~al.}(2016){Casey}, {Hogg}, {Ness}, {Rix}, {Ho}, \&
  {Gilmore}}]{Casey2016}
{Casey}, A.~R., {Hogg}, D.~W., {Ness}, M., {et~al.} 2016, ArXiv e-prints
  [\eprint[arXiv]{1603.03040}]

\bibitem[{{Cordiner} {et~al.}(2017){Cordiner}, {Cox}, {Lallement}, {Najarro},
  {Cami}, {Gull}, {Foing}, {Linnartz}, {Lindler}, {Proffitt}, {Sarre}, \&
  {Charnley}}]{Cordiner17}
{Cordiner}, M.~A., {Cox}, N.~L.~J., {Lallement}, R., {et~al.} 2017, \apjl, 843,
  L2

\bibitem[{{Cox} {et~al.}(2007){Cox}, {Boudin}, {Foing}, {Schnerr}, {Kaper},
  {Neiner}, {Henrichs}, {Donati}, \& {Ehrenfreund}}]{cox2007a}
{Cox}, N.~L.~J., {Boudin}, N., {Foing}, B.~H., {et~al.} 2007, \aap, 465, 899

\bibitem[{{Desert} {et~al.}(1995){Desert}, {Jenniskens}, \&
  {Dennefeld}}]{Desert1995}
{Desert}, F.-X., {Jenniskens}, P., \& {Dennefeld}, M. 1995, \aap, 303, 223

\bibitem[{{Eisenstein} {et~al.}(2011){Eisenstein}, {Weinberg}, {Agol},
  {Aihara}, {Allende Prieto}, {Anderson}, {Arns}, {Aubourg}, {Bailey},
  {Balbinot}, \& et~al.}]{Eisenstein11}
{Eisenstein}, D.~J., {Weinberg}, D.~H., {Agol}, E., {et~al.} 2011, AJ, 142, 72

\bibitem[{{Elyajouri} {et~al.}(2017{\natexlab{a}}){Elyajouri}, {Cox}, \&
  {Lallement}}]{Elyajouri2017b}
{Elyajouri}, M., {Cox}, N.~L.~J., \& {Lallement}, R. 2017{\natexlab{a}}, A\&A,
  605, L10

\bibitem[{{Elyajouri} {et~al.}(2018){Elyajouri}, {Lallement}, {Cox}, {Cami},
  {Cordiner}, {Smoker}, {Fahrang}, {Sarre}, \& {Linnartz}}]{Elyajouri2018}
{Elyajouri}, M., {Lallement}, R., {Cox}, N.~L.~J., {et~al.} 2018, \aap, 616,
  A143

\bibitem[{{Elyajouri} {et~al.}(2017{\natexlab{b}}){Elyajouri}, {Lallement},
  {Monreal-Ibero}, {Capitanio}, \& {Cox}}]{Elyajouri2017a}
{Elyajouri}, M., {Lallement}, R., {Monreal-Ibero}, A., {Capitanio}, L., \&
  {Cox}, N.~L.~J. 2017{\natexlab{b}}, A\&A, 600, A129

\bibitem[{{Elyajouri} {et~al.}(2016){Elyajouri}, {Monreal-Ibero}, {Remy}, \&
  {Lallement}}]{Elyajouri16}
{Elyajouri}, M., {Monreal-Ibero}, A., {Remy}, Q., \& {Lallement}, R. 2016,
  \apjs, 225, 19, (EMLR16)

\bibitem[{{Ensor} {et~al.}(2017){Ensor}, {Cami}, {Bhatt}, \&
  {Soddu}}]{Ensor2017}
{Ensor}, T., {Cami}, J., {Bhatt}, N.~H., \& {Soddu}, A. 2017, \apj, 836, 162

\bibitem[{{Fan} {et~al.}(2017){Fan}, {Welty}, {York}, {Sonnentrucker},
  {Dahlstrom}, {Baskes}, {Friedman}, {Hobbs}, {Jiang}, {Rachford}, {Snow},
  {Sherman}, \& {Zhao}}]{Fan2017}
{Fan}, H., {Welty}, D.~E., {York}, D.~G., {et~al.} 2017, \apj, 850, 194

\bibitem[{{Foing} \& {Ehrenfreund}(1994)}]{Foing94}
{Foing}, B.~H. \& {Ehrenfreund}, P. 1994, Nature, 369, 296

\bibitem[{{Friedman} {et~al.}(2011){Friedman}, {York}, {McCall}, {Dahlstrom},
  {Sonnentrucker}, {Welty}, {Drosback}, {Hobbs}, {Rachford}, \&
  {Snow}}]{Friedman11}
{Friedman}, S.~D., {York}, D.~G., {McCall}, B.~J., {et~al.} 2011, ApJ, 727, 33

\bibitem[{{Gaia Collaboration} {et~al.}(2016){Gaia Collaboration}, {Brown},
  {Vallenari}, {Prusti}, {de Bruijne}, {Mignard}, {Drimmel}, {Babusiaux},
  {Bailer-Jones}, {Bastian}, \& et~al.}]{Gaia2016}
{Gaia Collaboration}, {Brown}, A.~G.~A., {Vallenari}, A., {et~al.} 2016, \aap,
  595, A2

\bibitem[{{Garc{\'{\i}}a P{\'e}rez} {et~al.}(2016){Garc{\'{\i}}a P{\'e}rez},
  {Allende Prieto}, {Holtzman}, {Shetrone}, {M{\'e}sz{\'a}ros}, {Bizyaev},
  {Carrera}, {Cunha}, {Garc{\'{\i}}a-Hern{\'a}ndez}, {Johnson}, {Majewski},
  {Nidever}, {Schiavon}, {Shane}, {Smith}, {Sobeck}, {Troup}, {Zamora},
  {Weinberg}, {Bovy}, {Eisenstein}, {Feuillet}, {Frinchaboy}, {Hayden},
  {Hearty}, {Nguyen}, {O'Connell}, {Pinsonneault}, {Wilson}, \&
  {Zasowski}}]{Garcia2016}
{Garc{\'{\i}}a P{\'e}rez}, A.~E., {Allende Prieto}, C., {Holtzman}, J.~A.,
  {et~al.} 2016, \aj, 151, 144

\bibitem[{{Heger}(1922)}]{Heger22}
{Heger}, M.~L. 1922, Lick Observatory Bulletin, 10, 141

\bibitem[{{Herbig}(1995)}]{Herbig95}
{Herbig}, G.~H. 1995, ARA\&A, 33, 19

\bibitem[{{Holtzman} {et~al.}(2015){Holtzman}, {Shetrone}, {Johnson}, {Allende
  Prieto}, {Anders}, {Andrews}, {Beers}, {Bizyaev}, {Blanton}, {Bovy},
  {Carrera}, {Chojnowski}, {Cunha}, {Eisenstein}, {Feuillet}, {Frinchaboy},
  {Galbraith-Frew}, {Garc{\'{\i}}a P{\'e}rez}, {Garc{\'{\i}}a-Hern{\'a}ndez},
  {Hasselquist}, {Hayden}, {Hearty}, {Ivans}, {Majewski}, {Martell},
  {Meszaros}, {Muna}, {Nidever}, {Nguyen}, {OConnell}, {Pan}, {Pinsonneault},
  {Robin}, {Schiavon}, {Shane}, {Sobeck}, {Smith}, {Troup}, {Weinberg},
  {Wilson}, {Wood-Vasey}, {Zamora}, \& {Zasowski}}]{Holtzman15}
{Holtzman}, J.~A., {Shetrone}, M., {Johnson}, J.~A., {et~al.} 2015, AJ, 150,
  148

\bibitem[{{Jones}(2016)}]{Jones2016}
{Jones}, A.~P. 2016, Royal Society Open Science, 3, 160223

\bibitem[{{Lallement} {et~al.}(2019){Lallement}, {Babusiaux}, {Vergely},
  {Katz}, {Arenou}, {Valette}, {Hottier}, \& {Capitanio}}]{Lallement2019}
{Lallement}, R., {Babusiaux}, C., {Vergely}, J., {et~al.} 2019, arXiv e-prints
  [\eprint[arXiv]{1902.04116}]

\bibitem[{{Lallement} {et~al.}(2018){Lallement}, {Cox}, {Cami}, {Smoker},
  {Fahrang}, {Elyajouri}, {Cordiner}, {Linnartz}, {Smith}, {Ehrenfreund}, \&
  {Foing}}]{Lallement2018a}
{Lallement}, R., {Cox}, N.~L.~J., {Cami}, J., {et~al.} 2018, \aap, 614, A28

\bibitem[{{Lan} {et~al.}(2015){Lan}, {M{\'e}nard}, \& {Zhu}}]{Lan2015}
{Lan}, T.-W., {M{\'e}nard}, B., \& {Zhu}, G. 2015, \mnras, 452, 3629

\bibitem[{{Lef{\`e}vre} {et~al.}(2014){Lef{\`e}vre}, {Pagani}, {Juvela},
  {Paladini}, {Lallement}, {Marshall}, {Andersen}, {Bacmann}, {McGehee},
  {Montier}, {Noriega-Crespo}, {Pelkonen}, {Ristorcelli}, \&
  {Steinacker}}]{Lefevre2014}
{Lef{\`e}vre}, C., {Pagani}, L., {Juvela}, M., {et~al.} 2014, \aap, 572, A20

\bibitem[{{Majewski} {et~al.}(2017){Majewski}, {Schiavon}, {Frinchaboy},
  {Allende Prieto}, {Barkhouser}, {Bizyaev}, {Blank}, {Brunner}, {Burton},
  {Carrera}, {Chojnowski}, {Cunha}, {Epstein}, {Fitzgerald}, {Garc{\'{\i}}a
  P{\'e}rez}, {Hearty}, {Henderson}, {Holtzman}, {Johnson}, {Lam}, {Lawler},
  {Maseman}, {M{\'e}sz{\'a}ros}, {Nelson}, {Nguyen}, {Nidever}, {Pinsonneault},
  {Shetrone}, {Smee}, {Smith}, {Stolberg}, {Skrutskie}, {Walker}, {Wilson},
  {Zasowski}, {Anders}, {Basu}, {Beland}, {Blanton}, {Bovy}, {Brownstein},
  {Carlberg}, {Chaplin}, {Chiappini}, {Eisenstein}, {Elsworth}, {Feuillet},
  {Fleming}, {Galbraith-Frew}, {Garc{\'{\i}}a}, {Garc{\'{\i}}a-Hern{\'a}ndez},
  {Gillespie}, {Girardi}, {Gunn}, {Hasselquist}, {Hayden}, {Hekker}, {Ivans},
  {Kinemuchi}, {Klaene}, {Mahadevan}, {Mathur}, {Mosser}, {Muna}, {Munn},
  {Nichol}, {O'Connell}, {Parejko}, {Robin}, {Rocha-Pinto}, {Schultheis},
  {Serenelli}, {Shane}, {Silva Aguirre}, {Sobeck}, {Thompson}, {Troup},
  {Weinberg}, \& {Zamora}}]{Majewski2017}
{Majewski}, S.~R., {Schiavon}, R.~P., {Frinchaboy}, P.~M., {et~al.} 2017, \aj,
  154, 94

\bibitem[{{McCall} \& {Griffin}(2013)}]{McCall13}
{McCall}, B.~J. \& {Griffin}, R.~E. 2013, in Proceedings of the royal society
  A, Vol. 469, Proceedings of the royal society A, ed. {M.~Berry}, 20120604

\bibitem[{{Meyer} \& {Ulrich}(1984)}]{Meyer1984}
{Meyer}, D.~M. \& {Ulrich}, R.~K. 1984, \apj, 283, 98

\bibitem[{{Ness} {et~al.}(2015){Ness}, {Hogg}, {Rix}, {Ho}, \&
  {Zasowski}}]{Ness2015}
{Ness}, M., {Hogg}, D.~W., {Rix}, H.-W., {Ho}, A.~Y.~Q., \& {Zasowski}, G.
  2015, \apj, 808, 16

\bibitem[{{Planck Collaboration} {et~al.}(2016){Planck Collaboration}, {Adam},
  {Ade}, {Aghanim}, {Alves}, {Arnaud}, {Ashdown}, {Aumont}, {Baccigalupi},
  {Banday}, \& et~al.}]{Planck2016}
{Planck Collaboration}, {Adam}, R., {Ade}, P.~A.~R., {et~al.} 2016, \aap, 594,
  A10

\bibitem[{{Queiroz} {et~al.}(2018){Queiroz}, {Anders}, {Santiago}, {Chiappini},
  {Steinmetz}, {Dal Ponte}, {Stassun}, {da Costa}, {Maia}, {Crestani}, {Beers},
  {Fern{\'a}ndez-Trincado}, {Garc{\'{\i}}a-Hern{\'a}ndez}, {Roman-Lopes}, \&
  {Zamora}}]{Queiroz2018}
{Queiroz}, A.~B.~A., {Anders}, F., {Santiago}, B.~X., {et~al.} 2018, \mnras,
  476, 2556

\bibitem[{{Remy} {et~al.}(2018){Remy}, {Grenier}, {Marshall}, \&
  {Casandjian}}]{Remy2018}
{Remy}, Q., {Grenier}, I.~A., {Marshall}, D.~J., \& {Casandjian}, J.~M. 2018,
  \aap, 616, A71

\bibitem[{{Ruiterkamp} {et~al.}(2005){Ruiterkamp}, {Cox}, {Spaans}, {Kaper},
  {Foing}, {Salama}, \& {Ehrenfreund}}]{2005A&A...432..515R}
{Ruiterkamp}, R., {Cox}, N.~L.~J., {Spaans}, M., {et~al.} 2005, \aap, 432, 515

\bibitem[{{Santiago} {et~al.}(2016){Santiago}, {Brauer}, {Anders}, {Chiappini},
  {Queiroz}, {Girardi}, {Rocha-Pinto}, {Balbinot}, {da Costa}, {Maia},
  {Schultheis}, {Steinmetz}, {Miglio}, {Montalb{\'a}n}, {Schneider}, {Beers},
  {Frinchaboy}, {Lee}, \& {Zasowski}}]{Santiago2016}
{Santiago}, B.~X., {Brauer}, D.~E., {Anders}, F., {et~al.} 2016, \aap, 585, A42

\bibitem[{{Snow}(2014)}]{Snow2014}
{Snow}, T.~P. 2014, in IAU Symposium, Vol. 297, The Diffuse Interstellar Bands,
  ed. J.~{Cami} \& N.~L.~J. {Cox}, 3--12

\bibitem[{{Snow} \& {McCall}(2006)}]{SnowMcCall2006}
{Snow}, T.~P. \& {McCall}, B.~J. 2006, \araa, 44, 367

\bibitem[{{Snow} {et~al.}(2002){Snow}, {Welty}, {Thorburn}, {Hobbs}, {McCall},
  {Sonnentrucker}, \& {York}}]{Snow2002}
{Snow}, T.~P., {Welty}, D.~E., {Thorburn}, J., {et~al.} 2002, \apj, 573, 670

\bibitem[{{Snow} \& {Cohen}(1974)}]{Snow1974}
{Snow}, Jr., T.~P. \& {Cohen}, J.~G. 1974, \apj, 194, 313

\bibitem[{{Strom} {et~al.}(1975){Strom}, {Strom}, {Carrasco}, \&
  {Vrba}}]{Strom1975}
{Strom}, K.~M., {Strom}, S.~E., {Carrasco}, L., \& {Vrba}, F.~J. 1975, \apj,
  196, 489

\bibitem[{{Taylor}(2005)}]{Taylor2005}
{Taylor}, M.~B. 2005, in Astronomical Society of the Pacific Conference Series,
  Vol. 347, Astronomical Data Analysis Software and Systems XIV, ed.
  P.~{Shopbell}, M.~{Britton}, \& R.~{Ebert}, 29

\bibitem[{{Thorburn} {et~al.}(2003){Thorburn}, {Hobbs}, {McCall}, {Oka},
  {Welty}, {Friedman}, {Snow}, {Sonnentrucker}, \& {York}}]{Thorburn03}
{Thorburn}, J.~A., {Hobbs}, L.~M., {McCall}, B.~J., {et~al.} 2003, \apj, 584,
  339

\bibitem[{{Vuong} \& {Foing}(2000)}]{Vuong00}
{Vuong}, M.~H. \& {Foing}, B.~H. 2000, \aap, 363, L5

\bibitem[{{Wallerstein} \& {Cardelli}(1987)}]{Wallerstein1987}
{Wallerstein}, G. \& {Cardelli}, J.~A. 1987, \aj, 93, 1522

\bibitem[{{Wampler}(1966)}]{Wampler1966}
{Wampler}, E.~J. 1966, \apj, 144, 921

\bibitem[{{Xiang} {et~al.}(2017){Xiang}, {Li}, \& {Zhong}}]{Xiang2017}
{Xiang}, F.~Y., {Li}, A., \& {Zhong}, J.~X. 2017, \apj, 835, 107

\bibitem[{{Zasowski} {et~al.}(2013){Zasowski}, {Johnson}, {Frinchaboy},
  {Majewski}, {Nidever}, {Rocha Pinto}, {Girardi}, {Andrews}, {Chojnowski},
  {Cudworth}, {Jackson}, {Munn}, {Skrutskie}, {Beaton}, {Blake}, {Covey},
  {Deshpande}, {Epstein}, {Fabbian}, {Fleming}, {Garcia Hernandez}, {Herrero},
  {Mahadevan}, {M{\'e}sz{\'a}ros}, {Schultheis}, {Sellgren}, {Terrien}, {van
  Saders}, {Allende Prieto}, {Bizyaev}, {Burton}, {Cunha}, {da Costa},
  {Hasselquist}, {Hearty}, {Holtzman}, {Garc{\'{\i}}a P{\'e}rez}, {Maia},
  {O'Connell}, {O'Donnell}, {Pinsonneault}, {Santiago}, {Schiavon}, {Shetrone},
  {Smith}, \& {Wilson}}]{Zasowski13}
{Zasowski}, G., {Johnson}, J.~A., {Frinchaboy}, P.~M., {et~al.} 2013, AJ, 146,
  81

\bibitem[{{Zasowski} {et~al.}(2015){Zasowski}, {M{\'e}nard}, {Bizyaev},
  {Garc{\'{\i}}a-Hern{\'a}ndez}, {Garc{\'{\i}}a P{\'e}rez}, {Hayden},
  {Holtzman}, {Johnson}, {Kinemuchi}, {Majewski}, {Nidever}, {Shetrone}, \&
  {Wilson}}]{Zasowski15}
{Zasowski}, G., {M{\'e}nard}, B., {Bizyaev}, D., {et~al.} 2015, ApJ, 798, 35

\end{thebibliography}
\appendix

\section{Derivation of DIB EWs in the case of very small absorptions (illustrative examples)}
\clearpage
\begin{figure}
\centering
\includegraphics[width=0.5\textwidth]{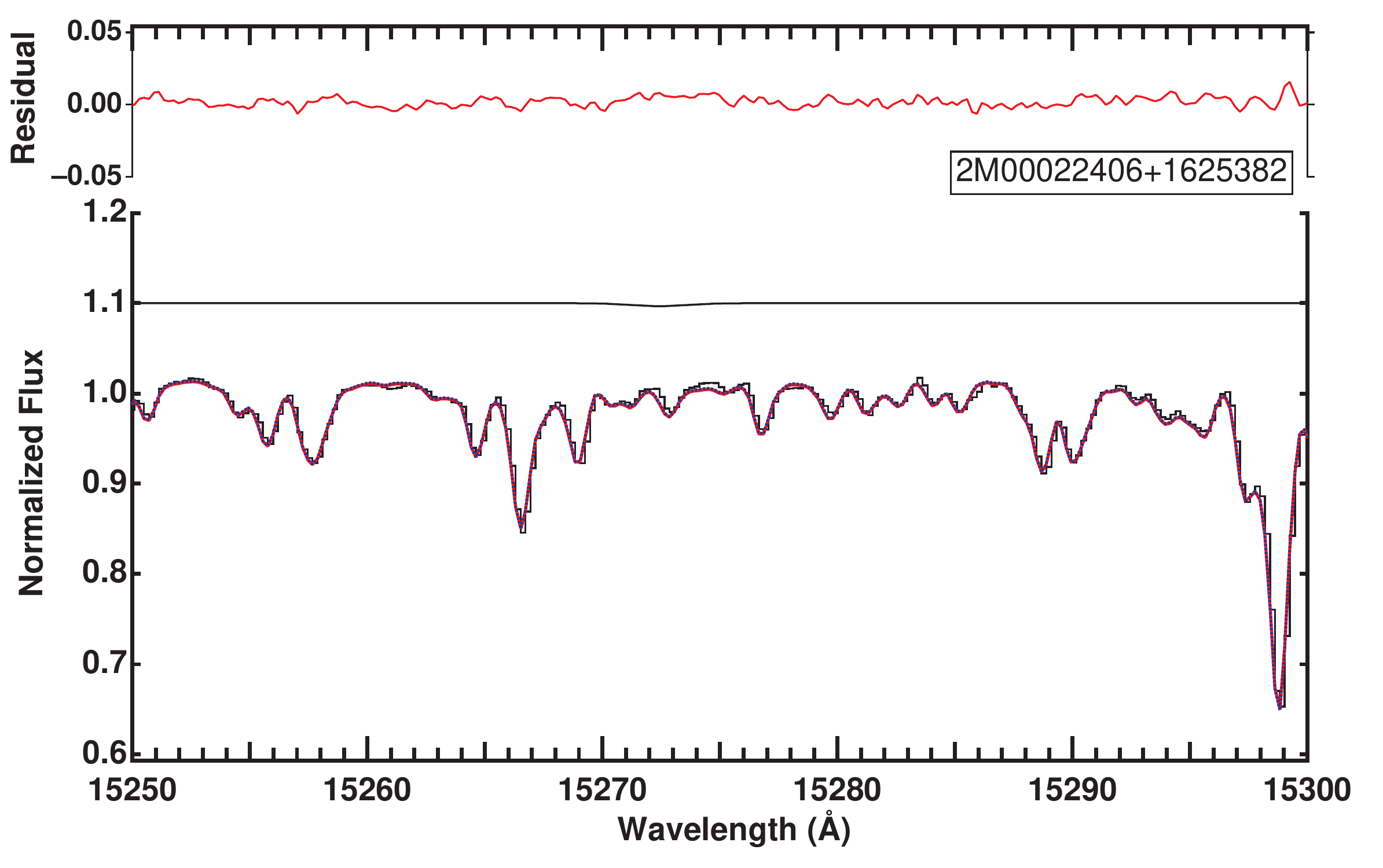}
\includegraphics[width=0.5\textwidth]{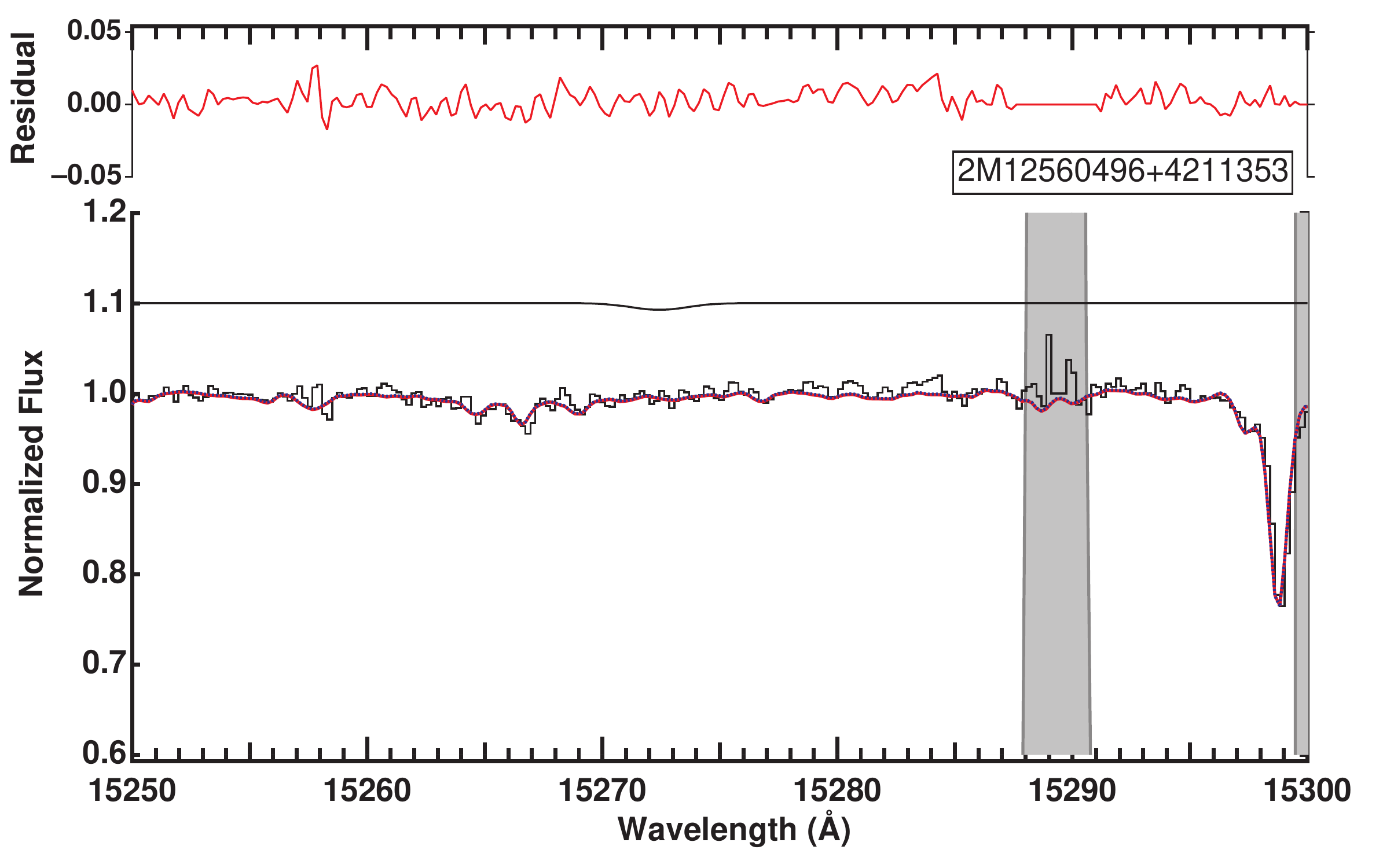}
\includegraphics[width=0.5\textwidth]{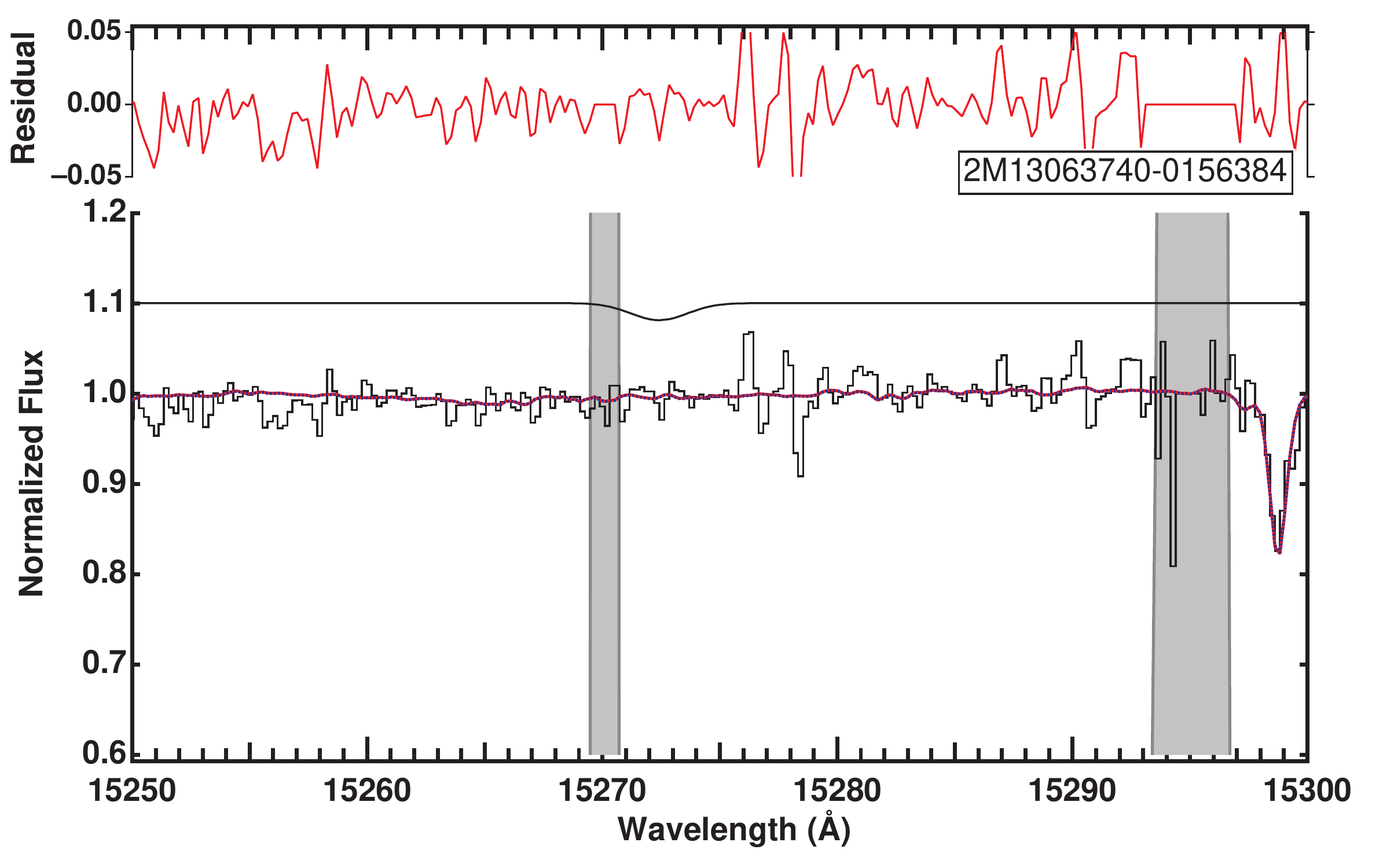}
  \caption{Illustration of EW upper limit derivation, in the case of non convergence of the Gaussian fit (see section 3.2). A Gaussian DIB profile with the maximum equivalent width compatible with the standard deviation in the data-stellar model residuals is shown at an imposed wavelength of  15272.42 $\mbox{\AA}$ in solid black (with an offset).}
    \label{fig:upperlimit_caselin}
\end{figure}
\begin{figure}
\centering
\includegraphics[width=0.5\textwidth]{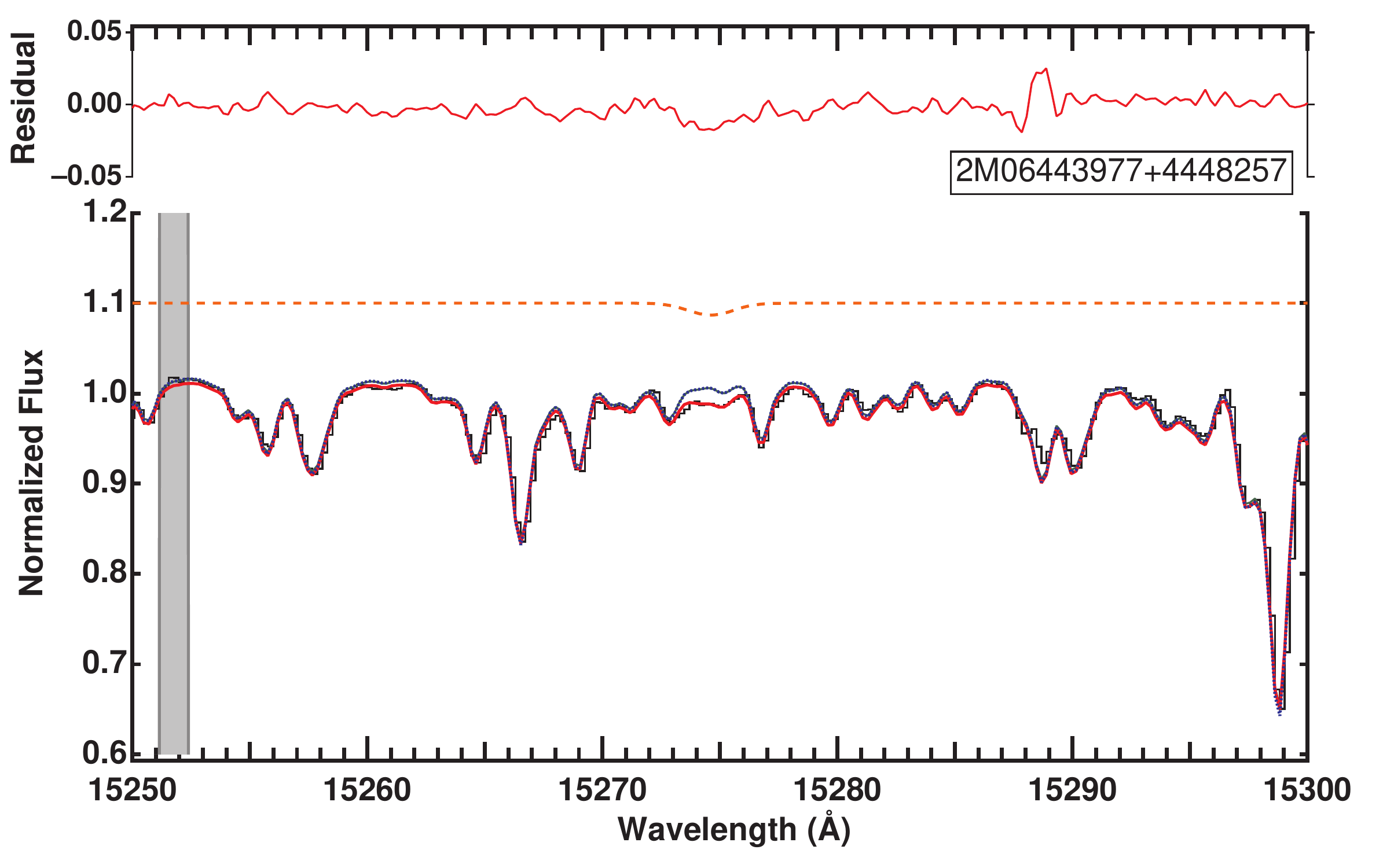}
\includegraphics[width=0.5\textwidth]{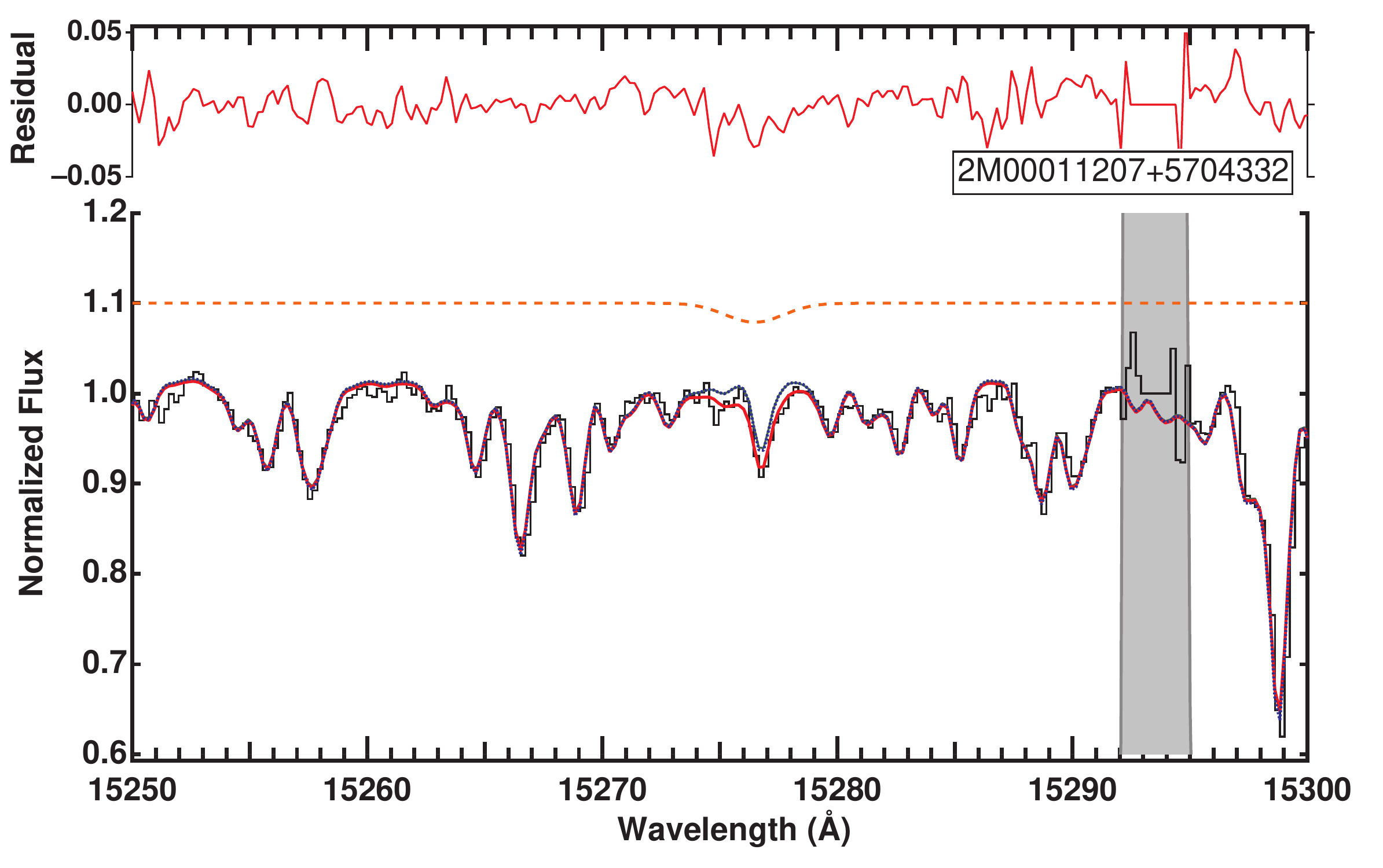}
\includegraphics[width=0.5\textwidth]{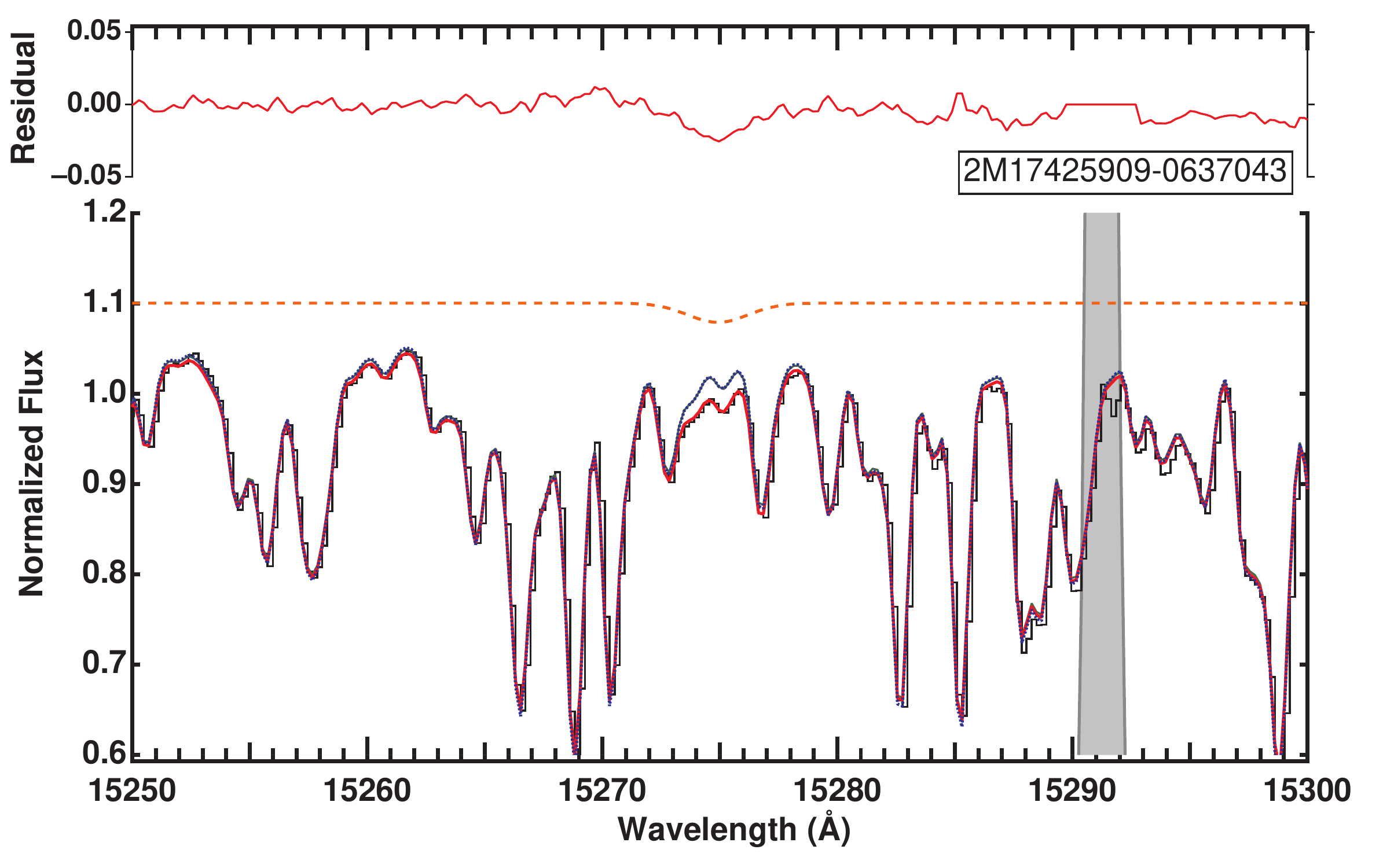}
  \caption{Illustration of DIB extraction in the case of convergence after derivation of an optimal DIB prior (the initial guess based on the preliminary computation of data-stellar model residuals and its Gaussian fit). The fitted DIB is shown in dashed orange with an offset.}
    \label{fig:upperlimit_case2}
\end{figure}
\begin{figure*}
    \centering
    \includegraphics[width=0.9\textwidth]{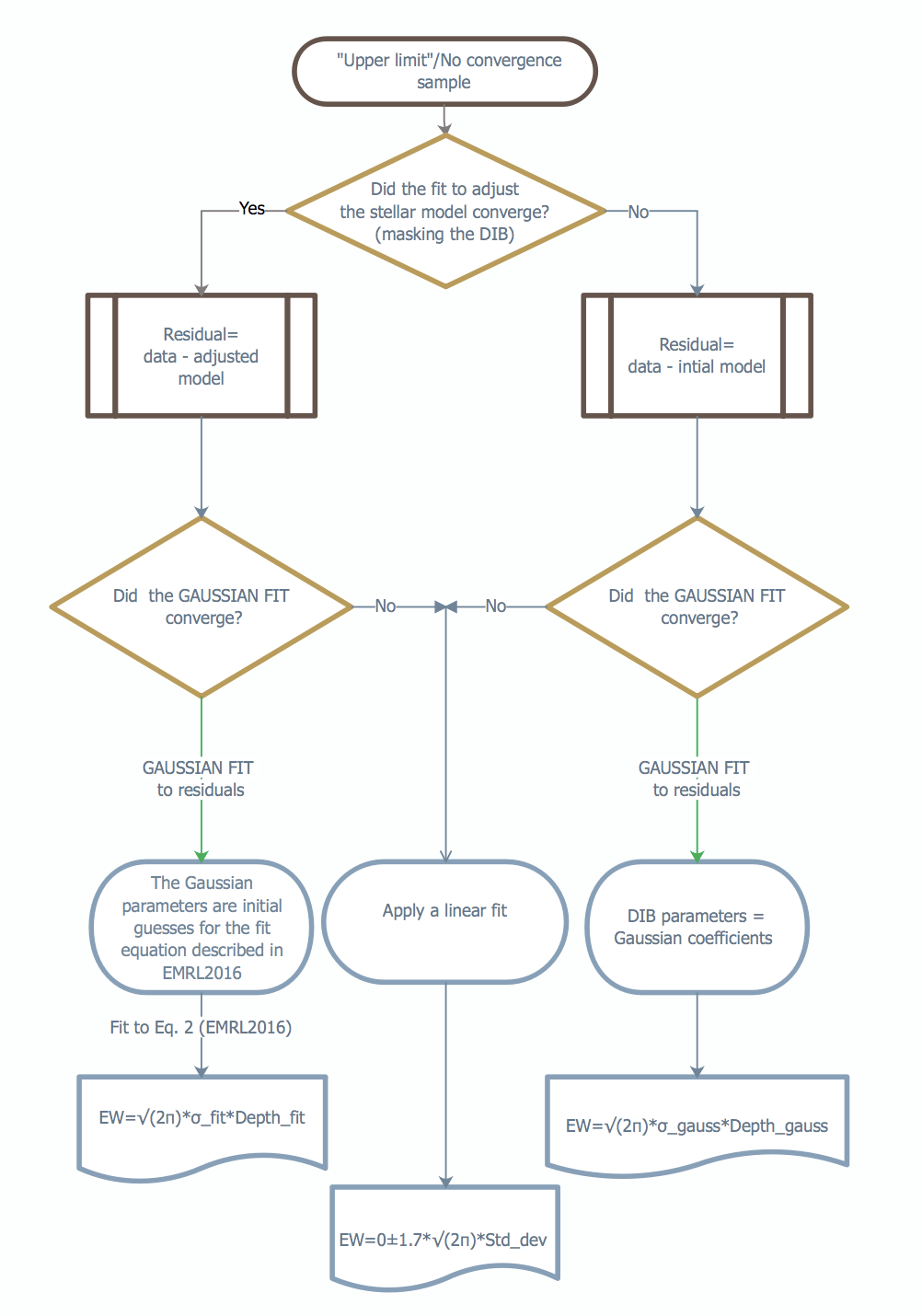}
    \caption{Flowchart compiling our decision criteria to determine the DIB equivalent widths in the upper limit and no convergence cases. Part of the spectra analyzed here are those classified as upper limit in our automated method described in \citetalias{Elyajouri16}. The width $\sigma_{fit}$ and the Depth depth$_{fit}$ are the Gaussian parameters derived from the same automated method described in \citetalias{Elyajouri16}. Std$_{dev}$ is the standard deviation derived from the linear fit to residuals over the wavelength region defined to measure the EW of the DIB.}
    \label{fig:flowchart_upper}
\end{figure*}
\end{document}